\documentclass[ reprint, amsmath,amssymb,aps, pra,superscriptaddress]{revtex4-2}

\usepackage{graphicx}
\usepackage{dcolumn}
\usepackage{bm}
\usepackage{soul}

\usepackage{braket}
\usepackage{graphicx}
\usepackage{amsmath,verbatim,latexsym,amssymb,indentfirst,mathrsfs,mathtools,amsthm,bbm,bm,hyperref,url}

\newcommand{\Tr}{{\rm Tr}}  

\newcommand*{\ketbra}[2]{\lvert #1 \rangle\!\langle #2 \rvert}

\newcommand*{\expval}[1]{\left\langle  #1  \right\rangle}
\usepackage{color}
\usepackage[dvipsnames]{xcolor}
\usepackage[normalem]{ulem}
\usepackage{physics}
\newcommand{\kater}[1]{{\color{black}#1}}
\newcommand{\kk}[1]{{\color{black}#1}}
\newcommand{\km}[1]{{\color{black}#1}}
\newcommand{\xingrui}[1]{{\color{black}#1}}

\newcommand{\mn}[1]{{\color{black}#1}}

\begin{document}

\title{Quantum process inference for a single qubit Maxwell's demon}

\author{Xingrui Song}
\affiliation{
Department of Physics, Washington University, St. Louis, Missouri 63130, USA
}
\author{Mahdi Naghiloo}
\affiliation{
Department of Physics, Washington University, St. Louis, Missouri 63130, USA
}
\affiliation{Research Laboratory of Electronics, MIT,  Cambridge, Massachusetts 02139}
\author{Kater Murch}
\affiliation{
Department of Physics, Washington University, St. Louis, Missouri 63130, USA
}

\date{\today}

\begin{abstract}
While quantum measurement theories are built around density matrices and observables, the laws of thermodynamics are based on processes such as \kk{are used in} heat engines and refrigerators.
The study of quantum thermodynamics fuses these two distinct paradigms. In this article, we highlight the usage of quantum process matrices as a unified language for describing thermodynamic processes in the quantum regime. 
\mn{We experimentally demonstrate this in the context of a quantum Maxwell’s demon, where two major quantities are commonly investigated; the average work extraction $\langle W \rangle$ and the efficacy $\gamma$ which measures how efficiently the feedback operation uses the obtained information}.
Using the tool of quantum process matrices, we \kk{develop} the optimal feedback protocols for these two quantities and experimentally investigate them in a superconducting circuit QED setup.
\end{abstract}

\maketitle

\section{introduction}

The interplay of information and energy is at the heart of \kk{thermodynamics,} originating from the thought experiment of Maxwell's demon \cite{Brillouin1956, Maruyama2009, leff2014maxwell, Parrondo2015, Lutz2015, Groenewold1971}. In particular, the laws of thermodynamics have been generalized to accommodate the presence of feedback operations  \cite{Sagawa2008, Ponmurugan2010, Sagawa2012, Lahiri2012, Abreu2012, Potts2018, Funo2013, Potts2019}. Experimental implementations of various types of classical demons have been realized \cite{Raizen2009, Serreli2007, Toyabe2010, Roldan2014, Koski2014, Koski2014-PNAS, Vidrighin2016}. The modern development of quantum technologies further enables us to investigate the idea of Maxwell's demon in the quantum regime \cite{Camati2016, Ciampini2017, Cottet2017, Masuyama2018, Annby-Andersson2020, Manzano2020, Naghiloo2018, Najera-Santos2020, Sanchez2019, Kumar2018}, \kk{where concepts such as coherence, entanglement, measurement backaction, and the exponential scaling of system Hilbert spaces may become important.  Furthermore, quantum information theory allows us to analyze and optimize these measurement and feedback-based protocols in a way that can reveal quantum thermodynamical advantages.}  

In this \km{article}, we introduce the tool of the quantum process matrix to analyze and optimize a weak-measurement-based Maxwell's demon protocol \cite{Naghiloo2018}. The quantum process matrix has vast application in quantum information processing \cite{Chow2011, Martinac2010, Rosenblum2018, Riebe2006, Yamamoto2010, Childs2001} and quantum optics \cite{Kupchak2015, Kim2018, Altepeter2003}, \kk{but its usage in quantum thermodynamics is still nascent \cite{Camati2016}. The optimization of feedback protocols has been considered in classical \cite{Horowitz2011} and quantum \cite{Manzano2018} contexts, with experimental implementation so far limited to classical systems \cite{Paneru2018}.   Using quantum process matrices, we are able to assign new meaning to the efficacy---a measure of how efficiently feedback uses obtained information \cite{Toyabe2010}---which can be related to violations of Jarzynski's equality when the role of information is neglected. Previous experimental work has demonstrated efficacy above unity \cite{Naghiloo2018}. However, optimization and maximization of the efficacy reveal certain fundamental limitations associated with the usual language of quantum mechanics.} First, while quantum mechanics provides us with methods to describe states and observables, thermodynamics concerns work, which is not an observable \cite{Talkner2007}. Second, we show that the quantum state alone---the density matrix---does not provide the full description of the evolution, and is thus inadequate for certain feedback tasks.   \kk{As a consequence, in order to design a feedback protocol that maximizes the efficacy, we harness the quantum process matrix to derive effective states that achieve this goal. Using a circuit QED setup, we experimentally test work and efficacy maximizing feedback protocols \xingrui{that utilize the quantum coherence encoded in the off-diagonal elements during the evolution. We} examine their performance over the parameter space of time, temperature, and measurement efficiency.} 

\km{This article is organizes as follows: In Section II we introduce the stochastic master equation that is used to track the quantum state of a qubit undergoing weak continuous measurement. We extend this stochastic master equation treatment  to derive a stochastic differential equation for the quantum process matrix that contains complete information about the quantum evolution. In Section III we introduce the protocol for a single qubit quantum Maxwell's demon along with the Jarzynski equality and the efficacy. We consider the optimization of feedback protocols that maximize different moments of the work distribution and study the performance of these protocols versus measurement efficiency and temperature. Several appendices discuss the experimental setup and data acquisition, the formalism of quantum process inference, analysis of the efficacy, the equivalence of different work distributions, methods used to reduce the measurement efficiency, statistical analysis methods, and the tomographic validation of quantum trajectories.}

\section{Continuous measurement of a superconducting qubit}

We use superconducting transmon qubit \cite{Paik2011,Koch2007} as a versatile platform for weak measurements and quantum state tracking. By coupling the qubit with a microwave cavity in the dispersive regime \cite{Wallraff2004, Hatridge2013, Murch2013, Weber2014, Gambetta2008}, one can perform continuous weak measurement and qubit state tracking without completely destroying coherences in the measurement basis. Using the measurement records, we experimentally reconstruct the time-dependent quantum process matrix along a single trajectory. 

\begin{figure*}
    \includegraphics{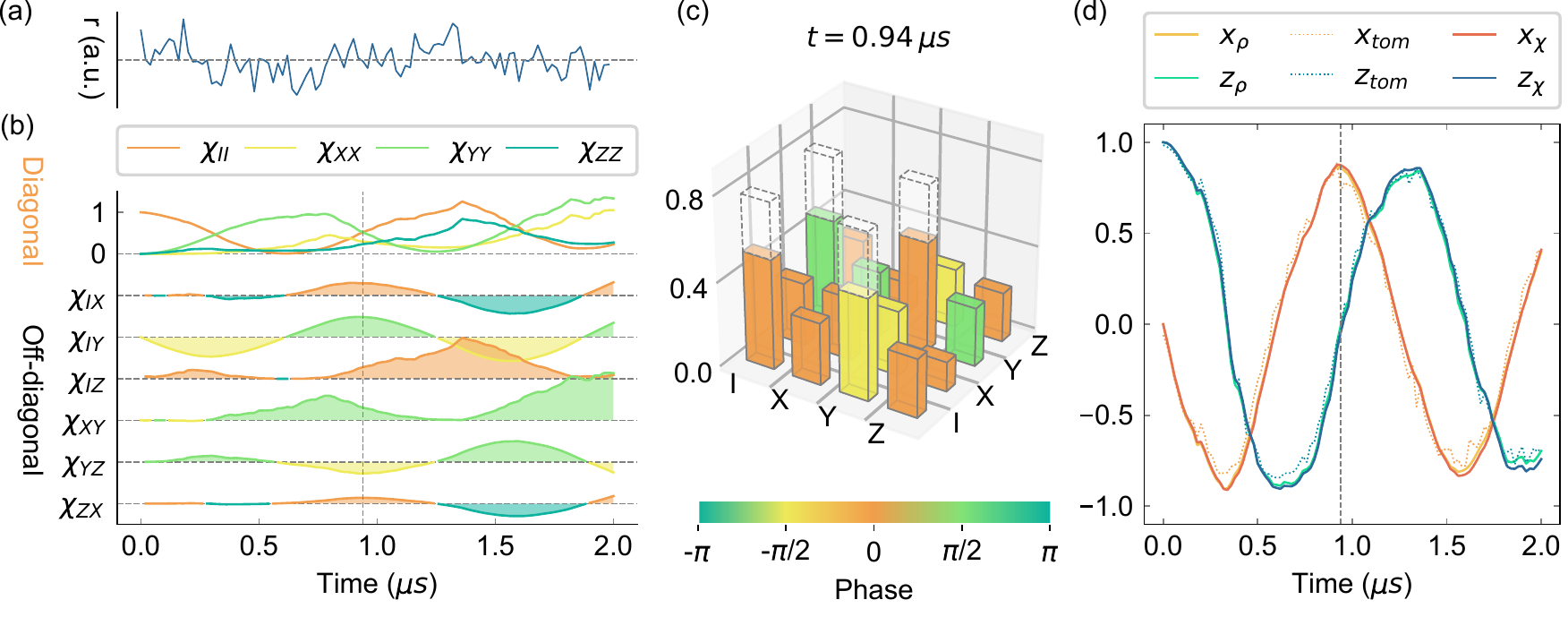}
    \caption{
        Quantum process matrix and quantum trajectory descriptions of the qubit evolution. (a) The signal $r$ obtained from continuous weak measurement. (b) The diagonal and off-diagonal matrix elements of the quantum process matrix $\chi_{jk}$. \xingrui{The colors of the shaded areas represent the phases of the off-diagonal elements.} (c) The matrix elements of the quantum process matrix at $t = 0.94 \: \mu$s. The length and color of each bar represent the norm and phase of the corresponding matrix elements of $\chi_{jk}$, respectively. The quantum process matrix of a (rescaled) ideal Hadamard gate (dashed boxes). (d) The quantum trajectory for $\rho_r$ calculated from a weak measurement record with the quantum process matrix ($x_\chi,z_\chi$) and with the SME ($x_\rho,z_\rho$) (solid lines). These trajectories are verified with tomographic validation ($x_\mathrm{tom},z_\mathrm{tom}$)  (dashed lines). In this example, the qubit is initialized in the ground state.
    }
    \label{fig:quantumproc}
\end{figure*}

The system is subject to a resonant drive given by the Hamiltonian $H_R = -\Omega_R\sigma_y/2$ in the rotating frame ($\Omega_R$ is the Rabi frequency and $\sigma_x,\  \sigma_y,\ \sigma_z$ are the Pauli operators, with $\sigma_z$ diagonal in the energy basis). The drive creates coherences between the qubit energy levels. Simultaneously, a continuous weak measurement probe signal \xingrui{coupled to $\sigma_z$} is used by the demon to track the state. The weak measurement record is denoted by $r(t)$ and the resulting conditional state evolution $\rho_r$ can be obtained from the stochastic master equation (SME) \cite{Naghiloo2018, Tan2015, Foroozani2016},
\begin{equation}
    \begin{aligned}
        \dot{\rho}_r = & \frac{1}{i\hbar}[H_R, \rho_r] + k(\sigma_z \rho_r \sigma_z - \rho_r) \\
        & + 2\eta k[\sigma_z \rho_r + \rho_r\sigma_z - 2 \Tr(\sigma_z \rho_r)\rho_r] r(t),
    \end{aligned}
    \label{SME}
\end{equation}
where $\eta$ is the efficiency of the detector and $k$ represents the strength of the measurement. In this measurement architecture, the signal $r(t)$ is the demodulated quadrature amplitude that encodes qubit state information (Fig.~\ref{fig:quantumproc}a), such that 
\begin{equation}
    r(t) = \langle \sigma_z \rangle(t) + d\mathcal{W},
\end{equation}
where $d\mathcal{W}$ is a zero-mean Gaussian distributed Wiener increment \cite{Jacobs2006}. This noise arises from the quantum fluctuations of the cavity probe. The noise obscures state information, resulting in weak measurement. 

We now introduce the tool of the quantum process matrix in our experiment, which represents the complete set of the information obtained from the measurement record \cite{Nielsen2010}. The evolution of the density matrix under the quantum operation $\mathcal{E}_r$ can be written as
\begin{equation}
    \mathcal{E}_r(\rho_i) = \sum_{jk} \chi_{jk}(r) K_j^{\dag} \rho_i K_k,
    \label{rhoeq}
\end{equation}
where $\rho_i$ is the initial density matrix of the system and $\chi_{jk}(r)$ are the elements of the quantum process matrix written in the basis of standard quantum process tomography (Fig.~\ref{fig:quantumproc}b,c)
\begin{equation}
    \left\lbrace K_j\right\rbrace = \left\lbrace I, \sigma_x, \sigma_y, \sigma_z\right\rbrace.
\end{equation}

Note that quantum operations are not trace-preserving. The resulting normalized density matrix is given by $\rho_r = \mathcal{E}_r(\rho_i) / \Tr \mathcal{E}_r(\rho_i)$. While there exists a technique of quantum process tomography to determine a quantum operation \cite{Nielsen2010}, it fails to apply to a time-dependent quantum process matrix, as we study here. This quantum process matrix is determined by a single stochastic measurement record, where repeated measurement and statistical averaging is impossible. Here, we develop an alternative way to infer the conditional quantum process by using a stochastic differential equation for the quantum process matrix (see Appendix B),
\begin{equation}
    \dot\chi_{jk}(r) = \sum_{mn}\sum_{m'n'}c_{mm'}^{j}c_{nn'}^{*k}\theta_{mn}(r)\chi_{m'n'}(r) ,
    \label{chieq}
\end{equation}
where $c_{jk}^{l}$ are the structure constants of the basis $\left\lbrace K_j \right\rbrace$ defined by
\begin{equation}\label{DefStructrueConstants}
    K_j K_k = \sum_{l} c_{jk}^{l}K_l ,
\end{equation}
whose complex conjugates are denoted by $c_{jk}^{*l}$. The coefficients $\theta_{mn}(r)$ are stochastic variables determined by the SME of the system. In our experimental setup, we have the closed form
\begin{equation}
    \theta(r) = 
    \begin{pmatrix}
        -k & 0 & -i\frac{\Omega_R}{2} & 2\eta kr \\
        0 & 0 & 0 & 0 \\
        i\frac{\Omega_R}{2} & 0 & k & 0 \\
        2\eta kr & 0 & 0 & 0
    \end{pmatrix}.
\end{equation}

Fig.~\ref{fig:quantumproc}b shows the evolution of the quantum process matrix obtained from one example measurement record. Clearly, the quantum process matrix contains more information about the system evolution than is preserved in the trajectories. $\chi(r)$ is a Hermitian matrix with positive diagonal elements. Although the information encoded in $\chi_{jk}(r)$ is generally obscure, in our experimental setup when the system undergoes $3/4$ of a Rabi cycle (approximately at $t = 0.94 \: \mu$s), the quantum operation can be compared with an ideal Hadamard gate where the effects of measurement backaction and dephasing are neglected (Fig.~\ref{fig:quantumproc}c). We also show that the quantum trajectory of the system can be recovered from $\chi(r)$. The result is compared with the trajectory expressed as Pauli expectation values $x = \mathrm{Tr}(\rho \sigma_x), z = \mathrm{Tr}(\rho \sigma_z)$, where $\rho$ is calculated from Eq.~\ref{SME} $(x_\rho, z_\rho)$ or via Eq.~\ref{rhoeq} ($x_\chi, z_\chi$). The trajectories generated from these two methods are nearly identical and agree with the tomographic validation (Fig.~\ref{fig:quantumproc}d). 


\section{Quantum thermodynamics}

\begin{figure}
    \centering
    \includegraphics{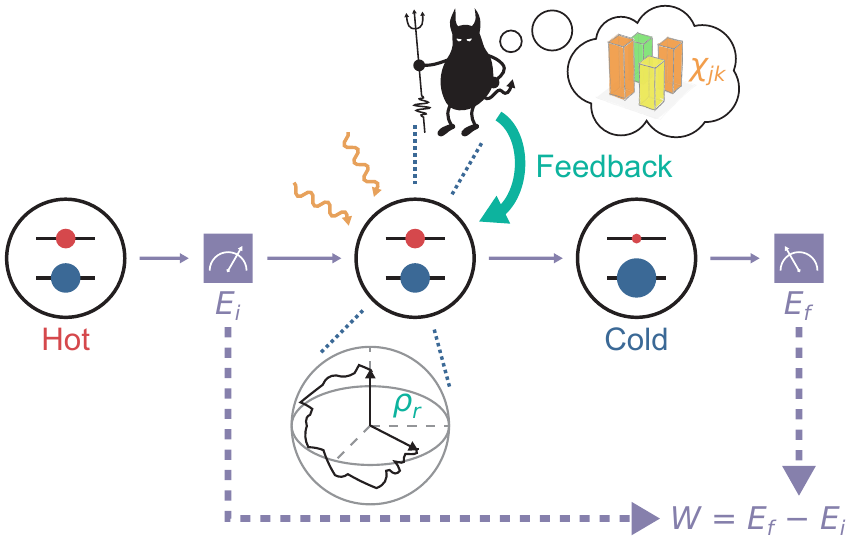}
    \caption{
        Demon utilizing quantum process inference. The qubit is prepared in a high-temperature initial state. We apply an initial measurement to determine $E_i$. The system undergoes Rabi drive and continuous weak measurement. The demon infers the quantum process matrix $\chi_{jk}$ from the measurement record and performs a feedback rotation to extract work. At the end, we apply a final measurement to determine $E_f$ concluding the TPM protocol. 
    }
    \label{fig:protocol}
\end{figure}

As is shown in Fig.~\ref{fig:protocol}, the qubit system is initialized \mn{in} the thermal state
\begin{equation}
  \rho_i=  \rho_{\mathrm{th}}(\beta) = \frac{1}{Z} e^{-\beta H} = \frac{1}{2\cosh(\beta / 2)}
    \begin{pmatrix}
    e^{\beta / 2} & 0 \\
    0 & e^{-\beta / 2}
    \end{pmatrix}.
\end{equation}
where $H = -\hbar\omega\sigma_z/2$ and $Z = 2 \cosh(\beta / 2)$ are the Hamiltonian and partition function of the qubit, respectively and $\beta = (k_B T)^{-1}$ is the inverse temperature. For simplicity, the qubit energy levels are given in units such that $\hbar\omega = 1$.

We consider a Maxwell's demon protocol where information from weak continuous measurement is used to extract work through unitary feedback. In order to experimentally determine the work extraction, we introduce the two-point measurement (TPM) protocol \cite{Campisi2011}, which consists of a pair of projective measurements at the beginning and the end of the measurement and feedback. The work extraction measured by the TPM protocol (Fig.~\ref{fig:protocol}) is calculated as
\begin{equation}
-W_{\mathrm{TPM}} = \expval{E_{i}} - \expval{E_{f}},
\end{equation}
where $E_i$ and $E_f$ represent the initial and final energy of the qubit system, respectively. Previous work \cite{Naghiloo2018} has studied a feedback protocol that was solely determined by the density matrix $\rho_r$. \km{Starting with a thermal state, the demon tracked the quantum trajectory of the qubit via the SME (Eq.~\ref{SME}). After a variable duration, the demon applied a feedback rotation to rotate the qubit state toward the ground state. This protocol maximized the work extracted from the qubit.}

\xingrui{In addition to $W$, the higher-order moments of the work distribution and their combinations are informative since they encode the correlation of the initial and final states of the system.  For example, the second order moment of $W$,
\begin{equation}
    \expval{W^2} = \expval{E_i^2} + \expval{E_f^2} - 2\expval{E_i E_f},
\end{equation}
explicitly involves the correlation between $E_i$ and $E_f$ in the cross term $\expval{E_i E_f}$, which, unlike a quantum mechanical observable, is inaccessible from the initial and the final density matrices of the system. With the language of quantum operations, this cross term can be expressed as
\begin{equation}
    \expval{E_i E_f} = \int \mathcal{D}r \Tr[H \mathcal{E}_r (\rho_{th}H)],
\end{equation}
where $\int \mathcal{D}r$ (see Appendix C) represents the path integral over the entire space of possible measurement records. In summary, the higher-order moments of the work distribution serve as the probe of the correlation information encoded in the quantum operation beyond a traditional density matrix treatment.

Among the various choices, one of the most valuable quantities to consider is given by the Jarzynski's equality \cite{Jarzynski1997, Jarzynski1997a}.} In the case that the initial and final free energies of the system are the same, Jarzynksi's equality is written,
\begin{equation}
    \label{Jar}
    \langle e^{-\beta W}\rangle = \gamma.
\end{equation}
 The equality introduces the efficacy, $\gamma$ \cite{Sagawa2010}. In the absence of measurement and feedback, $\gamma=1$. 
\km{ On one hand, $-W > 0$ is thermodynamic evidence of the demon extracting work while the $\gamma > 1$ is, on the other hand, more of an information-theoretical measure, since (i) $\gamma$ is dimensionless, and (ii) as we will show, $\gamma$ is bounded by $n$, the size of the system Hilbert space and is irrelevant to the energy spectrum of the system considered, } \xingrui{and (iii) $\gamma$ can only be maximized if the correlation information contained in the quantum process matrix is not destoyed.}

We now analyze the feedback protocols that maximize the work extraction and efficacy. Previous work has pointed out that the role of $\gamma$ is closely related to the idea of backward processes \cite{Sagawa2010}. Quantum operations can be utilized to study the time reversal of open quantum systems \cite{Crooks2008}. Motivated by these results, we also notice that the flexibility provided by quantum operations allows us to extract the information solely encoded in the measurement records without a specific initial state, which is of key importance for optimizing the feedback protocols considered in this letter. \xingrui{We emphasize that while the measurement records preserve the correlation between the initial and final states of the system, specifying an intial state in the SME Eq.~\eqref{SME} or Eq.~\eqref{rhoeq} may be invasive as this correlation information can be destroyed.} However, by replacing the initial state $\rho_i$ in Eq.~\eqref{rhoeq} with the completely mixed state $I / 2$, \xingrui{which can be understood as a ``least invasive'' choice,} we effectively discard any prior thermodynamic information about the system. The resulting quantity
\begin{figure}
    \centering
    \includegraphics{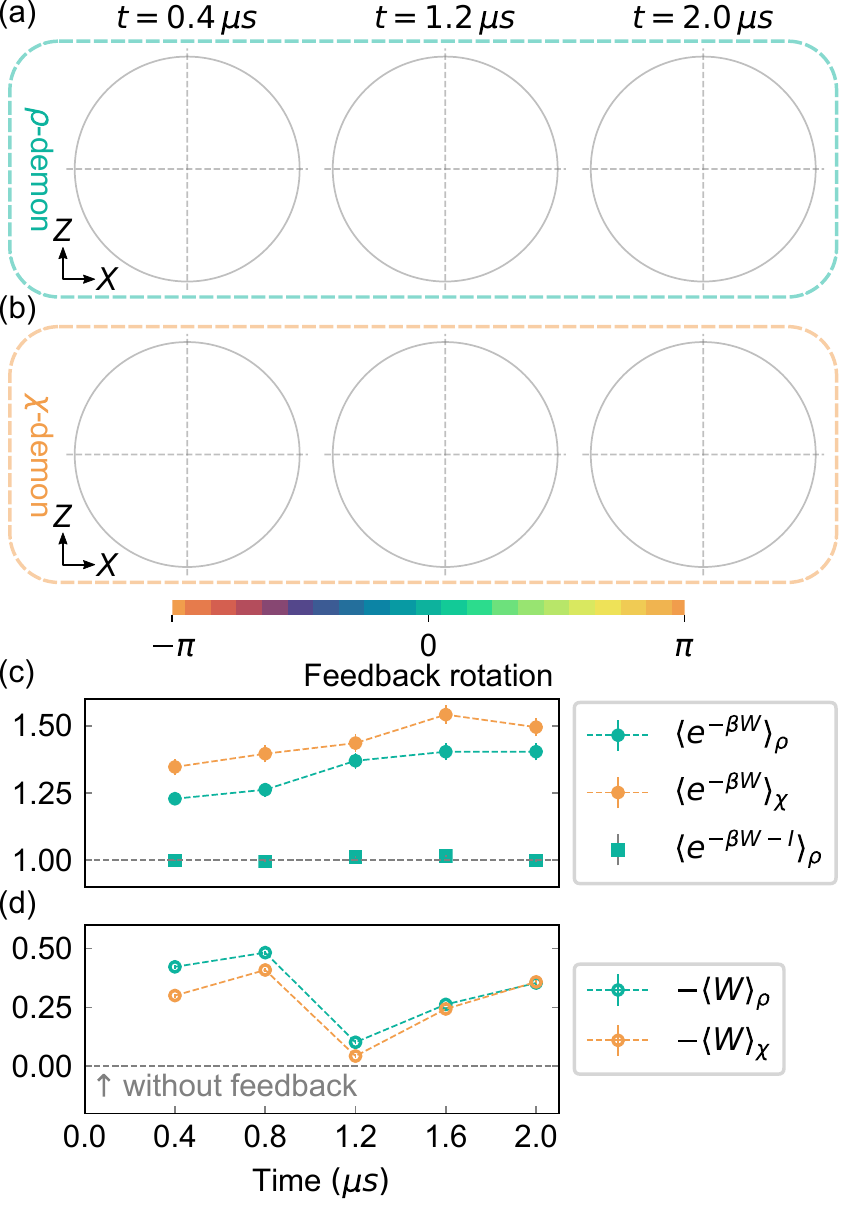}
    \caption{
        Two different feedback protocols. (a) The feedback protocol that maximizes the work extraction: an ensemble of $\rho_r$, obtained by tracking single trajectories via the SME from an initial thermal state $(\beta=1.3)$ are represented as points in  the $X$--$Z$ plane. As is shown in the plot, the feedback rotation angle is directly related to $\arctan(z / x)$ returning the state to the $+Z$-direction. Experimentally, this feedback rotation is approximated by 20 discrete angles separated by $\pi/10$. (b) The feedback protocol that maximizes the efficacy: the map is more complicated because the optimal feedback rotation (color) is determined by $\tilde{\rho}_r$ rather than $\rho_r$ (points on the $X$--$Z$ plane). (c) The efficacy of the two feedback protocols measured at different evolution times (round marks) for $\beta=2.5$. The generalized Jarzynski equality is verified for the $\rho$-demon (square marks). (d) Work advantage due to the two feedback protocols for $\beta=2.5$. 
    }
    \label{fig:feedback}
\end{figure}
\begin{equation}
    \tilde{\rho}_r \propto \mathcal{E}_r(I / 2) = \sum_{jk} \chi_{jk}(r) K_j^{\dag} K_k = \sum_{jkl} \chi_{jk}(r) c_{jk}^{l} K_l
\end{equation}
is the effective density matrix that we define, which is similarly normalized as $\tilde{\rho}_r = \mathcal{E}_r(I) / \Tr \mathcal{E}_r(I)$. The significance of $\tilde{\rho}_r$ becomes clearer if we rewrite the efficacy as
\begin{equation}
    \label{gamma_overlap}
    \gamma = \int \mathcal{D}r \, \Tr \left[ \rho_{\mathrm{th}}\mathcal{E}_r (I) \right] = n\mathbb{E}[\Tr(\rho_{\mathrm{th}} \tilde{\rho}_r)|I],
\end{equation}
where $n=2$ is the number of the energy levels of the system (see Appendix C). The expectation value $\mathbb{E}[\cdot|I]$ is evaluated as if the system were initialized into a completely mixed state. We comment that in the case where no feedback operation is performed on the system, the average of $\tilde{\rho}_r$ remains the completely mixed state $I / n$ and the value of Eq.~\eqref{gamma_overlap} reduces to 1 (Eq.~\eqref{Jar}). This equation also shows that $\gamma$ is proportional to the overlap between $\rho_\mathrm{th}$ and $\tilde{\rho}_r$. Noting the overlap $\Tr(\rho_{\mathrm{th}} \tilde{\rho}_r)$ never exceeds unity, we conclude that the efficacy of any feedback protocol is bounded by $n$, which can be exponentially large for multi-qubit systems.


After a unitary feedback $U_r$, the effective density matrix of the system becomes $\tilde{\rho}_f = U^{\dag}_r \tilde{\rho}_r U_r$. Since $\rho_{\mathrm{th}}$ has more population in the ground state, we conclude that an optimal feedback that maximizes the efficacy given $\tilde{\rho}_r$ will \kk{maximize} its overlap with $\rho_{\mathrm{th}}$ by returning $\tilde{\rho}_f$ to the $+Z$-direction, which defines the behavior of the ``$\chi$-demon'' in our experiment.  On the other hand, since in general, $\tilde{\rho}_r$ is not a function of $\rho_r$ and is obtained from $\chi(r)$, a ``$\rho$-demon'' unaware of $\tilde{\rho}_r$ is unable to perform this optimal feedback. This is the direct consequence of the fact that work is not an observable \cite{Talkner2007}.

\kk{As is shown in Fig.~\ref{fig:feedback}a,b} the feedback protocols designed for these two different tasks have very different behavior for the same ensemble of measurement records. Remarkably, we observe that maximizing the work extraction and maximizing the efficacy are generally incompatible tasks.

Both of the feedback protocols are able to achieve the $\gamma>1$ regime, as is shown in \kk{Fig.~\ref{fig:feedback}c}.  Especially, we confirm that the $\chi$-demon possesses a significant advantage over the $\rho$-demon under this measure. This advantage is larger for small $t$ because in the limit of long time evolution where the significance of $\rho_i$ decreases, we expect the difference between these two feedback protocols to vanish. \kk{Figure~\ref{fig:feedback}c}  also shows the generalized Jarzynski equality $\langle e^{-\beta W - \mathcal{I}} \rangle = 1$. This generalized equality accounts for the information exchange, defined as $\mathcal{I}_{i, f} = \ln P_{f}(\rho_r) - \ln P_{i}(\rho_i)$, where $P_i$ and $P_f$ represent the populations of the system before and after the evolution calculated in the instantaneous eigenbasis, respectively \cite{Naghiloo2018, Sagawa2008}.  \kk{Fig.~\ref{fig:feedback}d displays the extracted work due to feedback for the two protocols. As is expected, the $\rho$-demon}, which is optimized for work extraction performs better than the $\chi$-demon.

\begin{figure}
    \centering
    \includegraphics{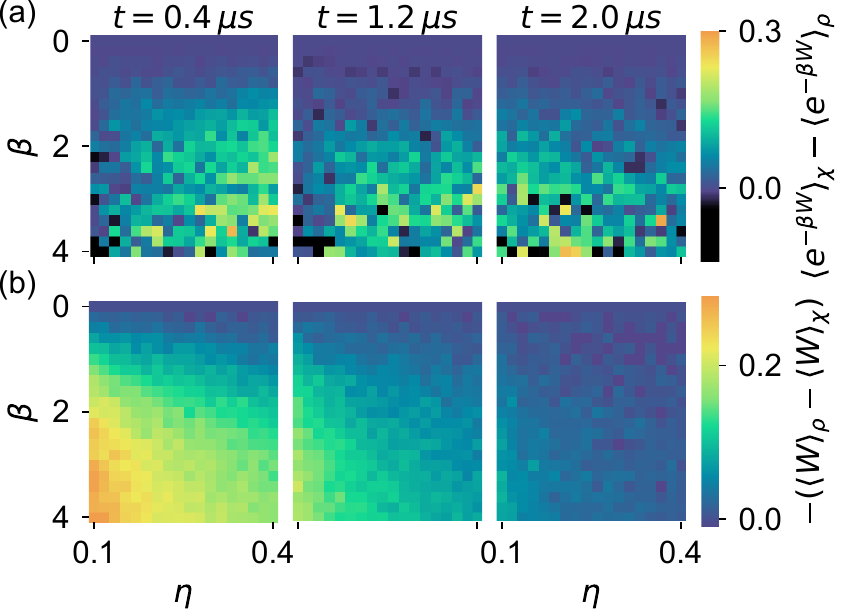}
    \caption{
        \kk{Efficacy and work extraction advantages of the two protocols. (a) The efficacy gain for different initial temperatures and different quantum efficiencies at three evolution times. \xingrui{The negative values (black) are caused by statistical fluctuations. Efficiency lower than the experimental realization is obtained by adding Gaussian random numbers as discussed in Appendix F}. (b) The corresponding work gain for the same parameters.}
    }
    \label{fig:betaeta}
\end{figure}

\kk{In order to build deeper intuition on the optimization of the two protocols, we examine the efficacy advantage ($\langle e^{-\beta W}\rangle_\chi-\langle e^{-\beta W}\rangle_\rho$) of the $\chi$-demon's protocol for different temperatures and measurement efficiencies.  Experimentally, we reduce the quantum efficiency by adding zero-mean Gaussian noise to our measurement signals, and different temperatures are obtained by sampling experiments with initial states given by respective Gibbs distributions. These data are displayed in Fig.~\ref{fig:betaeta}a. We first note that the $\chi$-demon always outperforms the $\rho$-demon in efficiently using the obtained information. This difference becomes most stark at short times and low temperature, where the  $\rho$-demon's feedback protocol is most significantly biased by the initial state.}
\kk{Fig.~\ref{fig:betaeta}b displays the corresponding work advantage ($-(\langle W \rangle_\rho - \langle W\rangle_\chi)$) of the $\rho$-demon. In regions of short evolution time, low temperature, and low quantum efficiency, initial state information is very relevant to work extraction. Likewise in the limits of high temperature, high efficiency, and long evolution time, the final state becomes less correlated with the initial state, leading to similar performance of the two protocols.}

\section{Outlook}

\kk{Experiments in quantum thermodynamics strive to elucidate opportunities for quantum advantage in thermodynamics, clarifying the interplay of measurement, information, and energy. We highlight the limitations of the quantum state alone for the optimization of feedback protocols, which can be addressed through the use of the quantum process matrix, which we track through continuous-time weak measurement.}  This work enables us to consider and optimize a broader variety of feedback protocols that take advantage of the information that is inaccessible in the density matrix alone, enabling new opportunities for achieving quantum thermodynamical advantages.

\begin{acknowledgements}
\kk{Acknowledgments---We thank J. Anders, S. Kizhakkumpurath, E. Lutz, and A. Romito for discussions. This research was supported by NSF Grant No. PHY-1752844 (CAREER) and used the facilities at the Institute of Materials Science and Engineering at Washington University. This project began at the KITP’s 2018 “Quantum Thermodynamics” conference and so was supported in part by the National Science Foundation under Grant No. NSF PHY-1748958.}
\end{acknowledgements}


	
	
	



\maketitle

\section*{Appendix A: Experimental setup}

The experimental setup in this work is identical to that used in reference \cite{Naghiloo2018}. Briefly, the system consists of a Transmon circuit ($E_C/h = 325$ MHz, $E_J/h = 8.88$ GHz, where $h$ is Planck's constant), embedded in a three dimensional microwave cavity ($\omega_c/2\pi = 6.86$ GHz). A dispersive interaction, characterized by a Hamiltonian term $\chi a^\dagger a \sigma_z$, with $\chi/2\pi = -0.3$ MHz, and $a^\dagger a$ the cavity number operator, leads to a qubit-state-dependent phase shift on a cavity probe. The weak cavity probe is amplified by a Josephson parametric amplifier operating in phase-sensitive mode, achieving an overall measurement quantum efficiency of $\eta = 0.48$. The experimental sequence consists of a strong (projective) measurement of the qubit energy, followed by variable duration evolution under continuous measurement with strength $k/2\pi=57$ kHz and $\Omega_R/2\pi = 0.8$ MHz, a feedback rotation, and finally a second projective energy measurement. The projective energy measurements are used for the TPM work distributions. The feedback operation is applied in a post-processing step; the data set contains different feedback rotations, and the subset of data where the correct rotations are chosen are selected from the data set for analysis. This allows for zero-latency feedback, especially when the computational overhead for calculating the quantum process matrix would require significant time. \xingrui{We treat the photons in the weak measurement probe signal as a free thermodynamic resource because the dispersive interaction only changes the phase of the incoming photons without changing their energy.}

\section*{Appendix B: Quantum process inference}

The stochastic master equation (SME) is written as
\begin{eqnarray}
    \dot{\rho_r} = -i \left[ H_R, \rho_r\right]  + k(\sigma_z \rho_r \sigma_z - \rho_r) \quad \quad \quad \nonumber \\\quad \quad \quad+ 2\eta k \left[ \sigma_z \rho_r + \rho_r \sigma_z -2 \Tr (\sigma_z \rho_r)\rho_r \right]r,
\end{eqnarray}
where $\rho(0) = \rho_i$ is the initial state.   Note that this equation is nonlinear in $\rho$ because of the quadratic term $\Tr (\sigma_z \rho)\rho$. In order to recover the linear nature of Kraus operators, we relax the restriction of trace \kater{preservation} to get
\begin{equation}
    \label{eqnphi}
    \dot{\varphi}_r = -i \left[ H_R, \varphi_r\right]  + k(\sigma_z \varphi_r \sigma_z - \varphi_r) + 2\eta k \left[ \sigma_z \varphi_r + \varphi_r \sigma_z \right]r,
\end{equation}
where $\varphi$ is the unnormalized density matrix with $\varphi(0) = \rho_i$. It can be verified that
\begin{equation}
    \rho_r = \frac{\varphi_r}{\Tr \varphi_r}.
\end{equation}
The right-hand-side of Eq.~\eqref{eqnphi} is represented in the basis $\{K_j\}$ as
\begin{equation}
    \label{phifromtheta}
    \dot{\varphi}_r = \sum_{j, k} \theta_{jk}(r) K_j \varphi_r K_k^{\dag},
\end{equation}
where the stochastic variables $\theta_{jk}$ are determined by the SME. Meanwhile, the evolution of $\varphi$ is also described by quantum operation $\mathcal{E}_{r}$ and the corresponding quantum process matrix $\chi$,
\begin{equation}
    \label{phifromchi}
    \varphi_r = \mathcal{E}_{r} (\rho_i) = \sum_{j, k} \chi_{jk}(r) K_j \rho_i K_k^{\dag}.
\end{equation}
\kater{Substituting} Eq.~\eqref{phifromchi} into Eq.\eqref{phifromtheta}, we obtain
\begin{equation}
\begin{aligned}
    & \sum_{j, k} \dot{\chi}_{jk}(r) K_j \rho_i K_k^{\dag} \\
    =& \sum_{m, n} \sum_{m', n'} \theta_{mn}(r) \chi_{m' n'}(r)K_m K_{m'}\rho_i K_{n'}^{\dag} K_n^{\dag} \\
    =& \sum_{m, n} \sum_{m', n'} c_{mm'}^{j} c_{nn'}^{*k}\theta_{mn}(r) \chi_{m' n'}(r)K_j \rho_i K_k^{\dag},
\end{aligned}
\end{equation}
where $c_{jk}^{l}$ are the structure constants of the basis $\{K_j\}$. By comparing the coefficients, we arrive at the stochastic differential equation for $\chi$
\begin{equation}
    \label{sdechi}
    \dot{\chi}_{jk}(r) = \sum_{m, n} \sum_{m', n'} c_{mm'}^{j} c_{nn'}^{*k}\theta_{mn}(r) \chi_{m' n'}(r).
\end{equation}
Using Eq. (\ref{sdechi}) and the initial quantum process matrix
\begin{equation}
    \chi_i = 
    \begin{pmatrix}
        1 & 0 & 0 & 0 \\
        0 & 0 & 0 & 0 \\
        0 & 0 & 0 & 0 \\
        0 & 0 & 0 & 0
    \end{pmatrix},
\end{equation}
we can therefore determine $\chi_{jk}(r)$. The structure constants $c^j_{mm'}$ form a 64-element tensor and are determined based on their definition in Eq.(\ref{DefStructrueConstants}).  We also note that because we have relaxed the trace preservation, the elements of $\chi(r)$ are statistically increasing in time, this growth in the elements of $\chi(r)$ can be seen in Fig.~1. This has no physical consequence because we explicitly normalize the density matrix determined by the quantum process.

\section*{Appendix C: Efficacy and the effective density matrix}

We study the statistical aspect of the formalism by first considering the operator-sum form of $\mathcal{E}_r$
\begin{equation}
    \mathcal{E}_r(\rho_i) = \sum_{j} M_{r,j}\rho_i M_{r,j}^{\dag},
\end{equation}
where $M_{r,j}$ is a set of Kraus operators implicitly determined by the $\chi$ matrix. With these Kraus operators, various types of probability density can be evaluated. Since each $M_{r,j}$ can be understood as an individual contribution to $\mathcal{E}_r$, by performing the summation over the traces,
\begin{equation}
    \label{eqnp}
    \sum_{j} \Tr \left( M_{r,j}\rho_i M_{r,j}^{\dag} \right) = \Tr  \mathcal{E}_{r} (\rho_i) = p(r|\rho_i), 
\end{equation}
we obtain the total probability density of getting the trajectory $r$ starting from initial state $\rho_i$ \cite{Nielsen2010}. Note that in Eq. (\ref{eqnp}), we have not incorporated the TPM protocol yet, which can be done by considering a particular pair of initial and final states $\ket{i}$ and $\ket{f}$.
By replacing the trace operation in Eq. (24) with the projection onto $\ket{f}$, we obtain
\begin{equation}
    \sum_{j} \mel{f}{M_{r,j}\rho_i M_{r,j}^{\dag}}{f} = \mel{f}{\mathcal{E}_{r} (\rho_i)}{f} = p(f, r|\rho_i).
\end{equation}
By further specifying the initial state to be $\ket{i}$, we obtain
\begin{equation}
    \sum_{j} \mel{f}{M_{r,j}\ketbra{i} M_{r,j}^{\dag}}{f} = \mel{f}{\mathcal{E}_{r} (\ketbra{i})}{f} = p(f, r| i).
\end{equation}
Then we arrive at the joint probability density $p(f, r, i)$ by writing
\begin{equation}
    \label{eqnpfri}
    p(f, r, i) = p(f, r| i) P_i = \mel{f}{\mathcal{E}_{r} (\ketbra{i})}{f} P_i,
\end{equation}
where $P_i$ represents the initial population of the system in $\ket{i}$. In this section, we will mainly focus on the two types of probability densities given by Eq. (\ref{eqnp}) and Eq. (\ref{eqnpfri}).

Noting that $r$ is a function of time, the normalization condition of these probability densities are properly expressed using the language of path \kater{integrals}, 
\begin{equation}
    \label{eqnDrp}
    \int \mathcal{D}r \, p(r|\rho_i) = 1,
\end{equation}
and
\begin{equation}
    \label{eqnDrp}
    \sum_{f, i} \int \mathcal{D}r \, p(f, r, i) = 1,
\end{equation}
respectively. Based on these normalization conditions, two types of expectation values over the ensemble of $r$ can be defined. For quantity $A(r, \rho_i)$ which explicitly depends on the initial density matrix, we define
\begin{equation}
\begin{aligned}
    \label{eqnDrprrhoi}
    \mathbb{E}[A|\rho_i] &= \int \mathcal{D}r \, p(r|\rho_i) A \\
    &= \int \mathcal{D}r \, A \Tr  \mathcal{E}_{r} (\rho_i),
\end{aligned}
\end{equation}
where the second line is obtained by applying Eq. (\ref{eqnp}). For quantity $A(f, r, i)$ which explicitly depends on the TPM measurement results, we define
\begin{equation}
    \label{eqnDrpri}
    \langle A \rangle = \sum_{f, i} \int \mathcal{D}r \, p(f, r, i) A.
\end{equation}
With these probabilities and expectation values properly defined, now we can use them to analyze the efficacy.

We consider an initial state described by the canonical ensemble at temperature $\beta$ with population in the $i$-th energy level given by
\begin{equation}
    \label{eqnPi}
    P_i(\beta) = \frac{1}{Z} e^{-\beta E_i}.
\end{equation}
The initial density matrix is described by the thermal state
\begin{equation}
    \label{rhoth}
    \rho_{\mathrm{th}} = \sum_{i} P_i(\beta) \ketbra{i}.
\end{equation}

Using Eq. (\ref{eqnDrpri}), the efficacy can be defined in a straightforward way as
\begin{equation}
    \label{gammaTPM}
    \gamma = \langle e^{-\beta W}\rangle = \sum_{f, i} \int \mathcal{D}r \, p(f, r, i) e^{-\beta(E_f - E_i)}.
\end{equation}
Eq. (\ref{gammaTPM}) can be simplified in several steps. By utilizing Eq. (\ref{eqnPi}), we can rewrite the exponential part to get
\begin{equation}
    \gamma = \sum_{f, i} \int \mathcal{D}r \, p(f, r, i) \frac{P_f(\beta)}{P_i(\beta)}.
\end{equation}
With Eq. (\ref{eqnpfri}) the probability part can be reformatted with $\mathcal{E}_r$
\begin{align}
    \gamma &= \sum_{f, i} \int \mathcal{D}r \,  \mel{f}{\mathcal{E}_{r} (\ketbra{i})}{f} P_i(\beta) \frac{P_f(\beta)}{P_i(\beta)} \nonumber \\
    &= \sum_{f, i} \int \mathcal{D}r \,  \mel{f}{\mathcal{E}_{r} (\ketbra{i})}{f} P_f(\beta).
\end{align}
Next, we deal with the summations. Since $\mathcal{E}_r$ is a linear mapping, the summation over $i$ is straightforward,
\begin{align}
    \gamma &= \sum_{f} \int \mathcal{D}r \,  \mel{f}{\mathcal{E}_{r} (\sum_i \ketbra{i})}{f} P_f(\beta) \nonumber\\ &= \sum_{f} \int \mathcal{D}r \,  \mel{f}{\mathcal{E}_{r} (I)}{f} P_f(\beta).
\end{align}
In order to perform the summation over $f$, we rewrite the quantum mechanical expectation value with the trace operation,
\begin{equation}
    \gamma = \int \mathcal{D}r \,  \Tr[\sum_{f} P_f(\beta)\ketbra{f} \mathcal{E}_{r} (I)].
\end{equation}
By utilizing Eq (\ref{rhoth}), we obtain,
\begin{equation}
    \label{gammaTr}
    \gamma = \int \mathcal{D}r \,  \Tr[\rho_{\mathrm{th}} \mathcal{E}_{r} (I)].
\end{equation}
The physical meaning of Eq. (\ref{gammaTr}) becomes clearer by introducing the effective density matrix
\begin{equation}
    \label{rhotilde}
    \tilde{\rho}_r = \frac{\mathcal{E}_r (I)}{\Tr \mathcal{E}_r (I)} = \frac{\mathcal{E}_r (I / n)}{\Tr \mathcal{E}_r (I / n)}.
\end{equation}
Combining Eq. (\ref{rhotilde}) and Eq. (\ref{gammaTr}), we obtain,
\begin{align}
    \gamma &= n \int \mathcal{D}r \, \Tr \left[ \rho_{\mathrm{th}}\mathcal{E}_{r} (I / n) \right] \nonumber \\ &= n \int \mathcal{D}r \, \Tr \left[ \rho_{\mathrm{th}}\tilde{\rho}_r \right] \Tr \mathcal{E}_r (I / n).
\end{align}
By comparing this result with Eq. (\ref{eqnDrprrhoi}), we arrive at
\begin{equation}
    \label{gammaE}
    \gamma = n\mathbb{E}[\Tr(\rho_{\mathrm{th}}\tilde{\rho}_r)|I],
\end{equation}
where the expectation value $\mathbb{E}[\cdot|I]$ is evaluated as if the system were initialized as a completely mixed state and the denominator $n$ has been omitted from the complete form $\mathbb{E}[\cdot|I / n]$ for convenience. Note that Eq. (\ref{gammaE}) is in the form of a nested expectation value because it is the statistical average of the quantum expectation value over the ensemble of measurement records.

\xingrui{
Similarly, we can evaluate
\begin{equation}
\begin{aligned}
    \expval{E_i E_f} &= \sum_{f, i} \int \mathcal{D}r \, p(f, r, i) E_i E_f \\
    &= \sum_{f, i} \int \mathcal{D}r \,  \mel{f}{\mathcal{E}_{r} (\ketbra{i})}{f} P_i(\beta) E_i E_f \\
    &= \sum_{f} \int \mathcal{D}r \,  \mel{f}{\mathcal{E}_{r} (\rho_{th} H)}{f} E_f \\
    &= \int \mathcal{D}r \Tr[H \mathcal{E}_r (\rho_{th}H)].
\end{aligned}
\end{equation}
}

\section*{Appendix D: Work distribution}
\begin{figure}[h]
    \centering
    \includegraphics{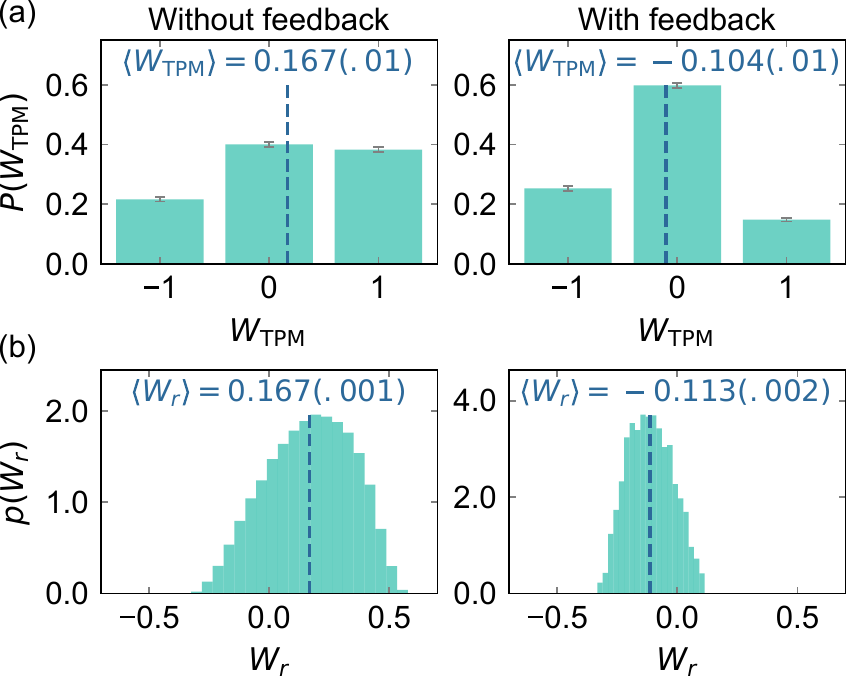}
    \caption{
        Two approaches to calculate the work distribution. The left and right panels display the work distribution before and after the feedback rotation, respectively. (a) The TPM work distribution measured at $\beta=0.5$. The dashed lines indicate the average work. (b) The corresponding conditional work distribution viewed by the demon. $\langle W_{r} \rangle$ is in fair agreement with $\langle W_{\mathrm{TPM}} \rangle$.}
    \label{fig:workdist}
\end{figure}

To characterize how the demon extracts work we compare the work distribution with and without unitary feedback. The work distribution obtained from the TPM is always discrete, but this quantity does not reflect the true expectation from the demon's point of view because the demon has no prior knowledge of the TPM measurement results, the actual work distribution viewed by the demon is the conditional expectation value
\begin{equation}
    \label{Wr}
    -W_{r} = - \mathbb{E}[W|r, \rho_{i}] = \Tr(\rho_{i} H) - \Tr(\rho_{f} H).
\end{equation}
The conditional work extraction depends on the stochastic measurement record $r$ and is a continuous variable taking values from $-1$ to $1$. We experimentally recover the conditional work extraction for each trajectory and obtain its statistical distribution (Fig.~\ref{fig:workdist}).

 Naturally, these two different descriptions of work produce the same overall expectation value, as is guaranteed by the law of total expectation,
\begin{equation}
    \langle W_{\mathrm{TPM}} \rangle =  \langle W_{r} \rangle,
\end{equation}
where $\langle W_{r} \rangle$ is defined from Eq. (\ref{eqnDrprrhoi}) as
\begin{equation}
    \langle W_{r} \rangle = \int \mathcal{D}r \, p(r|\rho_i) W_r.
\end{equation}

From Eq. (\ref{Wr}), we also see the optimal feedback  that  maximizes the work extraction will minimize the overlap between $\rho_f$ and $H$ by returning the state to the +$Z$-direction, which corresponds to the behavior of the ``$\rho$-demon'' in our experiment.

\xingrui{\section*{Appendix E: Optimization of the feedback protocols}

We optimize the feedback protocols by considering the expectation value of a given operator $A$ with respect to the density matrix $\rho_r$ under a unitary feedback operation $U$. The final state and the final expectation value after the feedback are written as
\begin{equation}
    \rho_{f} = U \rho_{r} U^{\dag},
\end{equation}
and
\begin{equation}
    \expval{A}_f = \Tr (\rho_f A),
\end{equation}
respectively. If $U$ is the optimal feedback, we expect $\expval{A}_f$ to stay unchanged under an arbitrary additional infinitesimal unitary operation
\begin{equation}
    V = I+iJ\delta\lambda,
\end{equation}
up to the first order in $\delta\lambda$, where $J$ is a Hermitian operator and $\delta\lambda$ is an infinitesimal real parameter.
By applying the infinitesimal operation $V$ to $\rho_f$, we obtain
\begin{equation}
\delta\rho_f=(I+iJ\delta\lambda)\rho_f(I-iJ\delta\lambda)-\rho_f=i[ J,\rho_f]\delta\lambda.
\end{equation}
The resulting variation of $\langle A \rangle_f$,
\begin{equation}
\delta\langle A \rangle_f=i\delta\lambda\Tr([ J,\rho_f]A).
\end{equation}
Since $J$ is arbitrary, we can choose $J$ to be a projection operator
\begin{equation}
J=\ket{\psi}\bra{\psi},
\end{equation}
where $\ket{\psi}$ is an arbitary pure state. Then $\langle A \rangle_f$ reduces to
\begin{equation}
\delta\langle A \rangle_f=i\mel{\psi}{[\rho_f,A]}{\psi}\delta\lambda.
\end{equation}
Note here $[\rho_f,A]$ is an anti-Hermitian operator. The condition $\delta\langle A \rangle_f=0$ for any $\ket{\psi}$ implies
\begin{equation}
[\rho_f,A]=0.
\end{equation}
In other words, the optimal feedback operation always makes the final density matrix diagonalized in the basis defined by $A$. As a consequence, for the $\rho$-demon which maximizes the work extraction, the optimal feedback operation diagonalizes the density matrix in the energy basis, with the larger population occupying the lower energy states. For the $\chi$-demon which maximizes the efficacy given by Eq. (\ref{gammaE}), the effective density matrix defined by Eq. (\ref{rhotilde}) is used, instead.

}

\section*{Appendix F: Lowering measurement efficiency with added Gaussian noise}

\xingrui{In Fig.~\ref{fig:betaeta},} we study the behavior of the feedback protocols at low measurement efficiency by adding zero-mean Gaussian random numbers into the measurement record while processing the data. The variance $\sigma^2$ of the noise is determined by the equality,
\begin{equation}
    \left(\frac{d\mathcal{W}}{\sqrt{4k\eta}\Delta t}\right)^2 + \sigma^2 = \left(\frac{d\mathcal{W}}{\sqrt{4k\eta'}\Delta t}\right)^2,
\end{equation}
where $\eta$ is the efficiency of experimental set up and $\eta'$ is the effective measurement efficiency with the added noise.

\section*{Appendix G: Statistical analysis}

The error analysis in Fig.~3 and Fig.~5 relies on the formula,
\begin{equation}
    \Delta A = \sqrt{\frac{\langle A^2\rangle - \langle A\rangle^2}{N}},
\end{equation}
where $A$ represents the quantity  averaged over and $N$ is the number of measurement records  used. Experimentally, the quantity $A$ and $A^2$ are determined for each measurement record and their mean values are calculated separately. In Fig.~3c,d, a total data set of 676,072 measurement records is used. Of this data set, we select subensembles that meet the specified feedback protocols with $N$ varying from 6,205 to 6,669. In Fig.~3d, the work advantage is displayed as $-(\langle W\rangle_{\rho,\chi} - \langle W\rangle_{n})$, where $\langle W\rangle_{n}$ represents the work extraction with no feedback. The corresponding statistical uncertainties are determined by error propagation as $\sqrt{(\Delta W_{\rho,\chi})^2 + (\Delta W_{n})^2}$. The statistical uncertainties of $\langle e^{-\beta W}\rangle$, $\langle e^{-\beta W - \mathcal{I}}\rangle$ and $-(\langle W\rangle_{\rho,\chi} - \langle W\rangle_{n})$ are displayed as vertical bars. In Fig.~5, a total of 68,856 measurement records is used, with subensembles of around 3,300. The errors of $P(W_{\mathrm{TPM}})$, $\langle W_{\mathrm{TPM}}\rangle$ and $\langle W_{r}\rangle$ are included in the corresponding figures.

\section*{Appendix H: Tomographic validation}

We validate the prediction of the quantum trajectories by performing quantum state tomography over a subensemble (Fig.~1d). We first generate a reference quantum trajectory from the measurement record shown in Fig.~1a. For each time $t$, the quantum trajectory predicts a pair of expectation values $ x(t) $ and $ z(t) $ (solid lines). This pair of expectation values are validated by preparing an ensemble of trajectories with an identical experimental setup but an evolution time truncated to $t$. Then we examine a subset of this ensemble such that their prediction on the final state is close enough to $ x(t) $ (or $ z(t) $), within $\pm 0.04$ tolerance. Note that although these trajectories may behave differently prior to $t$, ideally they share the common final expectation value. Since each of the trajectories \kk{are} followed by a final projective measurement, we are allowed to apply quantum state tomography to examine this subensemble. The resulting expectation values (dashed lines) given by the tomography are compared with the reference trajectory.                                                                                                                                                                                 


\begin{thebibliography}{60}%
\makeatletter
\providecommand \@ifxundefined [1]{%
 \@ifx{#1\undefined}
}%
\providecommand \@ifnum [1]{%
 \ifnum #1\expandafter \@firstoftwo
 \else \expandafter \@secondoftwo
 \fi
}%
\providecommand \@ifx [1]{%
 \ifx #1\expandafter \@firstoftwo
 \else \expandafter \@secondoftwo
 \fi
}%
\providecommand \natexlab [1]{#1}%
\providecommand \enquote  [1]{``#1''}%
\providecommand \bibnamefont  [1]{#1}%
\providecommand \bibfnamefont [1]{#1}%
\providecommand \citenamefont [1]{#1}%
\providecommand \href@noop [0]{\@secondoftwo}%
\providecommand \href [0]{\begingroup \@sanitize@url \@href}%
\providecommand \@href[1]{\@@startlink{#1}\@@href}%
\providecommand \@@href[1]{\endgroup#1\@@endlink}%
\providecommand \@sanitize@url [0]{\catcode `\\12\catcode `\$12\catcode
  `\&12\catcode `\#12\catcode `\^12\catcode `\_12\catcode `\%12\relax}%
\providecommand \@@startlink[1]{}%
\providecommand \@@endlink[0]{}%
\providecommand \url  [0]{\begingroup\@sanitize@url \@url }%
\providecommand \@url [1]{\endgroup\@href {#1}{\urlprefix }}%
\providecommand \urlprefix  [0]{URL }%
\providecommand \Eprint [0]{\href }%
\providecommand \doibase [0]{https://doi.org/}%
\providecommand \selectlanguage [0]{\@gobble}%
\providecommand \bibinfo  [0]{\@secondoftwo}%
\providecommand \bibfield  [0]{\@secondoftwo}%
\providecommand \translation [1]{[#1]}%
\providecommand \BibitemOpen [0]{}%
\providecommand \bibitemStop [0]{}%
\providecommand \bibitemNoStop [0]{.\EOS\space}%
\providecommand \EOS [0]{\spacefactor3000\relax}%
\providecommand \BibitemShut  [1]{\csname bibitem#1\endcsname}%
\let\auto@bib@innerbib\@empty
\bibitem [{\citenamefont {Brillouin}\ and\ \citenamefont
  {Hellwarth}(1956)}]{Brillouin1956}%
  \BibitemOpen
  \bibfield  {author} {\bibinfo {author} {\bibfnamefont {L.}~\bibnamefont
  {Brillouin}}\ and\ \bibinfo {author} {\bibfnamefont {R.~W.}\ \bibnamefont
  {Hellwarth}},\ }\bibfield  {title} {\bibinfo {title} {{ Science and
  Information Theory }},\ }\href {https://doi.org/10.1063/1.3059856} {\bibfield
   {journal} {\bibinfo  {journal} {Physics Today}\ }\textbf {\bibinfo {volume}
  {9}},\ \bibinfo {pages} {39} (\bibinfo {year} {1956})}\BibitemShut {NoStop}%
\bibitem [{\citenamefont {Maruyama}\ \emph {et~al.}(2009)\citenamefont
  {Maruyama}, \citenamefont {Nori},\ and\ \citenamefont
  {Vedral}}]{Maruyama2009}%
  \BibitemOpen
  \bibfield  {author} {\bibinfo {author} {\bibfnamefont {K.}~\bibnamefont
  {Maruyama}}, \bibinfo {author} {\bibfnamefont {F.}~\bibnamefont {Nori}},\
  and\ \bibinfo {author} {\bibfnamefont {V.}~\bibnamefont {Vedral}},\
  }\bibfield  {title} {\bibinfo {title} {{Colloquium: The physics of Maxwell's
  demon and information}},\ }\href {https://doi.org/10.1103/RevModPhys.81.1}
  {\bibfield  {journal} {\bibinfo  {journal} {Reviews of Modern Physics}\
  }\textbf {\bibinfo {volume} {81}},\ \bibinfo {pages} {1} (\bibinfo {year}
  {2009})},\ \Eprint {https://arxiv.org/abs/0707.3400} {arXiv:0707.3400}
  \BibitemShut {NoStop}%
\bibitem [{\citenamefont {Leff}\ and\ \citenamefont
  {Rex}(2014)}]{leff2014maxwell}%
  \BibitemOpen
  \bibfield  {author} {\bibinfo {author} {\bibfnamefont {H.~S.}\ \bibnamefont
  {Leff}}\ and\ \bibinfo {author} {\bibfnamefont {A.~F.}\ \bibnamefont {Rex}},\
  }\href
  {https://press.princeton.edu/books/hardcover/9780691634432/maxwells-demon}
  {\emph {\bibinfo {title} {{Maxwell's Demon: Entropy, Information,
  Computing}}}},\ Princeton Series in Physics\ (\bibinfo  {publisher}
  {Princeton University Press},\ \bibinfo {year} {2014})\BibitemShut {NoStop}%
\bibitem [{\citenamefont {Parrondo}\ \emph {et~al.}(2015)\citenamefont
  {Parrondo}, \citenamefont {Horowitz},\ and\ \citenamefont
  {Sagawa}}]{Parrondo2015}%
  \BibitemOpen
  \bibfield  {author} {\bibinfo {author} {\bibfnamefont {J.~M.}\ \bibnamefont
  {Parrondo}}, \bibinfo {author} {\bibfnamefont {J.~M.}\ \bibnamefont
  {Horowitz}},\ and\ \bibinfo {author} {\bibfnamefont {T.}~\bibnamefont
  {Sagawa}},\ }\bibfield  {title} {\bibinfo {title} {{Thermodynamics of
  information}},\ }\href {https://doi.org/10.1038/nphys3230} {\bibfield
  {journal} {\bibinfo  {journal} {Nature Physics}\ }\textbf {\bibinfo {volume}
  {11}},\ \bibinfo {pages} {131} (\bibinfo {year} {2015})}\BibitemShut
  {NoStop}%
\bibitem [{\citenamefont {Lutz}\ and\ \citenamefont
  {Ciliberto}(2015)}]{Lutz2015}%
  \BibitemOpen
  \bibfield  {author} {\bibinfo {author} {\bibfnamefont {E.}~\bibnamefont
  {Lutz}}\ and\ \bibinfo {author} {\bibfnamefont {S.}~\bibnamefont
  {Ciliberto}},\ }\bibfield  {title} {\bibinfo {title} {{Information: From
  Maxwell's demon to Landauer's eraser}},\ }\href
  {https://doi.org/10.1063/PT.3.2912} {\bibfield  {journal} {\bibinfo
  {journal} {Physics Today}\ }\textbf {\bibinfo {volume} {68}},\ \bibinfo
  {pages} {30} (\bibinfo {year} {2015})}\BibitemShut {NoStop}%
\bibitem [{\citenamefont {Groenewold}(1971)}]{Groenewold1971}%
  \BibitemOpen
  \bibfield  {author} {\bibinfo {author} {\bibfnamefont {H.~J.}\ \bibnamefont
  {Groenewold}},\ }\bibfield  {title} {\bibinfo {title} {{A problem of
  information gain by quantal measurements}},\ }\href
  {https://doi.org/10.1007/BF00815357} {\bibfield  {journal} {\bibinfo
  {journal} {International Journal of Theoretical Physics}\ }\textbf {\bibinfo
  {volume} {4}},\ \bibinfo {pages} {327} (\bibinfo {year} {1971})}\BibitemShut
  {NoStop}%
\bibitem [{\citenamefont {Sagawa}\ and\ \citenamefont
  {Ueda}(2008)}]{Sagawa2008}%
  \BibitemOpen
  \bibfield  {author} {\bibinfo {author} {\bibfnamefont {T.}~\bibnamefont
  {Sagawa}}\ and\ \bibinfo {author} {\bibfnamefont {M.}~\bibnamefont {Ueda}},\
  }\bibfield  {title} {\bibinfo {title} {{Second law of thermodynamics with
  discrete quantum feedback control}},\ }\href
  {https://doi.org/10.1103/PhysRevLett.100.080403} {\bibfield  {journal}
  {\bibinfo  {journal} {Physical Review Letters}\ }\textbf {\bibinfo {volume}
  {100}},\ \bibinfo {pages} {080403} (\bibinfo {year} {2008})},\ \Eprint
  {https://arxiv.org/abs/0710.0956} {arXiv:0710.0956} \BibitemShut {NoStop}%
\bibitem [{\citenamefont {Ponmurugan}(2010)}]{Ponmurugan2010}%
  \BibitemOpen
  \bibfield  {author} {\bibinfo {author} {\bibfnamefont {M.}~\bibnamefont
  {Ponmurugan}},\ }\bibfield  {title} {\bibinfo {title} {{Generalized detailed
  fluctuation theorem under nonequilibrium feedback control}},\ }\href
  {https://doi.org/10.1103/PhysRevE.82.031129} {\bibfield  {journal} {\bibinfo
  {journal} {Physical Review E - Statistical, Nonlinear, and Soft Matter
  Physics}\ }\textbf {\bibinfo {volume} {82}},\ \bibinfo {pages} {031129}
  (\bibinfo {year} {2010})},\ \Eprint {https://arxiv.org/abs/1005.4311}
  {arXiv:1005.4311} \BibitemShut {NoStop}%
\bibitem [{\citenamefont {Sagawa}(2012)}]{Sagawa2012}%
  \BibitemOpen
  \bibfield  {author} {\bibinfo {author} {\bibfnamefont {T.}~\bibnamefont
  {Sagawa}},\ }\bibfield  {title} {\bibinfo {title} {{Thermodynamics of
  information processing in small systems}},\ }\href
  {https://doi.org/10.1143/PTP.127.1} {\bibfield  {journal} {\bibinfo
  {journal} {Progress of Theoretical Physics}\ }\textbf {\bibinfo {volume}
  {127}},\ \bibinfo {pages} {1} (\bibinfo {year} {2012})}\BibitemShut {NoStop}%
\bibitem [{\citenamefont {Lahiri}\ \emph {et~al.}(2012)\citenamefont {Lahiri},
  \citenamefont {Rana},\ and\ \citenamefont {Jayannavar}}]{Lahiri2012}%
  \BibitemOpen
  \bibfield  {author} {\bibinfo {author} {\bibfnamefont {S.}~\bibnamefont
  {Lahiri}}, \bibinfo {author} {\bibfnamefont {S.}~\bibnamefont {Rana}},\ and\
  \bibinfo {author} {\bibfnamefont {A.~M.}\ \bibnamefont {Jayannavar}},\
  }\bibfield  {title} {\bibinfo {title} {{Fluctuation theorems in the presence
  of information gain and feedback}},\ }\href
  {https://doi.org/10.1088/1751-8113/45/6/065002} {\bibfield  {journal}
  {\bibinfo  {journal} {Journal of Physics A: Mathematical and Theoretical}\
  }\textbf {\bibinfo {volume} {45}},\ \bibinfo {pages} {065002} (\bibinfo
  {year} {2012})},\ \Eprint {https://arxiv.org/abs/1109.6508} {arXiv:1109.6508}
  \BibitemShut {NoStop}%
\bibitem [{\citenamefont {Abreu}\ and\ \citenamefont
  {Seifert}(2012)}]{Abreu2012}%
  \BibitemOpen
  \bibfield  {author} {\bibinfo {author} {\bibfnamefont {D.}~\bibnamefont
  {Abreu}}\ and\ \bibinfo {author} {\bibfnamefont {U.}~\bibnamefont
  {Seifert}},\ }\bibfield  {title} {\bibinfo {title} {{Thermodynamics of
  genuine nonequilibrium states under feedback control}},\ }\href
  {https://doi.org/10.1103/PhysRevLett.108.030601} {\bibfield  {journal}
  {\bibinfo  {journal} {Physical Review Letters}\ }\textbf {\bibinfo {volume}
  {108}},\ \bibinfo {pages} {030601} (\bibinfo {year} {2012})},\ \Eprint
  {https://arxiv.org/abs/1109.5892} {arXiv:1109.5892} \BibitemShut {NoStop}%
\bibitem [{\citenamefont {Potts}\ and\ \citenamefont
  {Samuelsson}(2018)}]{Potts2018}%
  \BibitemOpen
  \bibfield  {author} {\bibinfo {author} {\bibfnamefont {P.~P.}\ \bibnamefont
  {Potts}}\ and\ \bibinfo {author} {\bibfnamefont {P.}~\bibnamefont
  {Samuelsson}},\ }\bibfield  {title} {\bibinfo {title} {{Detailed Fluctuation
  Relation for Arbitrary Measurement and Feedback Schemes}},\ }\href
  {https://doi.org/10.1103/PhysRevLett.121.210603} {\bibfield  {journal}
  {\bibinfo  {journal} {Physical Review Letters}\ }\textbf {\bibinfo {volume}
  {121}},\ \bibinfo {pages} {210603} (\bibinfo {year} {2018})},\ \Eprint
  {https://arxiv.org/abs/1807.05034} {arXiv:1807.05034} \BibitemShut {NoStop}%
\bibitem [{\citenamefont {Funo}\ \emph {et~al.}(2013)\citenamefont {Funo},
  \citenamefont {Watanabe},\ and\ \citenamefont {Ueda}}]{Funo2013}%
  \BibitemOpen
  \bibfield  {author} {\bibinfo {author} {\bibfnamefont {K.}~\bibnamefont
  {Funo}}, \bibinfo {author} {\bibfnamefont {Y.}~\bibnamefont {Watanabe}},\
  and\ \bibinfo {author} {\bibfnamefont {M.}~\bibnamefont {Ueda}},\ }\bibfield
  {title} {\bibinfo {title} {{Integral quantum fluctuation theorems under
  measurement and feedback control}},\ }\href
  {https://doi.org/10.1103/PhysRevE.88.052121} {\bibfield  {journal} {\bibinfo
  {journal} {Physical Review E - Statistical, Nonlinear, and Soft Matter
  Physics}\ }\textbf {\bibinfo {volume} {88}},\ \bibinfo {pages} {052121}
  (\bibinfo {year} {2013})},\ \Eprint {https://arxiv.org/abs/1307.2362}
  {arXiv:1307.2362} \BibitemShut {NoStop}%
\bibitem [{\citenamefont {Potts}\ and\ \citenamefont
  {Samuelsson}(2019)}]{Potts2019}%
  \BibitemOpen
  \bibfield  {author} {\bibinfo {author} {\bibfnamefont {P.~P.}\ \bibnamefont
  {Potts}}\ and\ \bibinfo {author} {\bibfnamefont {P.}~\bibnamefont
  {Samuelsson}},\ }\bibfield  {title} {\bibinfo {title} {{Thermodynamic
  uncertainty relations including measurement and feedback}},\ }\href
  {https://doi.org/10.1103/PhysRevE.100.052137} {\bibfield  {journal} {\bibinfo
   {journal} {Physical Review E}\ }\textbf {\bibinfo {volume} {100}},\ \bibinfo
  {pages} {052137} (\bibinfo {year} {2019})},\ \Eprint
  {https://arxiv.org/abs/1904.04913} {arXiv:1904.04913} \BibitemShut {NoStop}%
\bibitem [{\citenamefont {Raizen}(2009)}]{Raizen2009}%
  \BibitemOpen
  \bibfield  {author} {\bibinfo {author} {\bibfnamefont {M.~G.}\ \bibnamefont
  {Raizen}},\ }\bibfield  {title} {\bibinfo {title} {{Comprehensive control of
  atomic motion}},\ }\href {https://doi.org/10.1126/science.1171506} {\bibfield
   {journal} {\bibinfo  {journal} {Science}\ }\textbf {\bibinfo {volume}
  {324}},\ \bibinfo {pages} {1403} (\bibinfo {year} {2009})}\BibitemShut
  {NoStop}%
\bibitem [{\citenamefont {Serreli}\ \emph {et~al.}(2007)\citenamefont
  {Serreli}, \citenamefont {Lee}, \citenamefont {Kay},\ and\ \citenamefont
  {Leigh}}]{Serreli2007}%
  \BibitemOpen
  \bibfield  {author} {\bibinfo {author} {\bibfnamefont {V.}~\bibnamefont
  {Serreli}}, \bibinfo {author} {\bibfnamefont {C.~F.}\ \bibnamefont {Lee}},
  \bibinfo {author} {\bibfnamefont {E.~R.}\ \bibnamefont {Kay}},\ and\ \bibinfo
  {author} {\bibfnamefont {D.~A.}\ \bibnamefont {Leigh}},\ }\bibfield  {title}
  {\bibinfo {title} {{A molecular information ratchet}},\ }\href
  {https://doi.org/10.1038/nature05452} {\bibfield  {journal} {\bibinfo
  {journal} {Nature}\ }\textbf {\bibinfo {volume} {445}},\ \bibinfo {pages}
  {523} (\bibinfo {year} {2007})}\BibitemShut {NoStop}%
\bibitem [{\citenamefont {Toyabe}\ \emph {et~al.}(2010)\citenamefont {Toyabe},
  \citenamefont {Sagawa}, \citenamefont {Ueda}, \citenamefont {Muneyuki},\ and\
  \citenamefont {Sano}}]{Toyabe2010}%
  \BibitemOpen
  \bibfield  {author} {\bibinfo {author} {\bibfnamefont {S.}~\bibnamefont
  {Toyabe}}, \bibinfo {author} {\bibfnamefont {T.}~\bibnamefont {Sagawa}},
  \bibinfo {author} {\bibfnamefont {M.}~\bibnamefont {Ueda}}, \bibinfo {author}
  {\bibfnamefont {E.}~\bibnamefont {Muneyuki}},\ and\ \bibinfo {author}
  {\bibfnamefont {M.}~\bibnamefont {Sano}},\ }\bibfield  {title} {\bibinfo
  {title} {{Experimental demonstration of information-to-energy conversion and
  validation of the generalized Jarzynski equality}},\ }\href
  {https://doi.org/10.1038/nphys1821} {\bibfield  {journal} {\bibinfo
  {journal} {Nature Physics}\ }\textbf {\bibinfo {volume} {6}},\ \bibinfo
  {pages} {988} (\bibinfo {year} {2010})}\BibitemShut {NoStop}%
\bibitem [{\citenamefont {Rold{\'{a}}n}\ \emph {et~al.}(2014)\citenamefont
  {Rold{\'{a}}n}, \citenamefont {Mart{\'{i}}nez}, \citenamefont {Parrondo},\
  and\ \citenamefont {Petrov}}]{Roldan2014}%
  \BibitemOpen
  \bibfield  {author} {\bibinfo {author} {\bibfnamefont {{\'{E}}.}~\bibnamefont
  {Rold{\'{a}}n}}, \bibinfo {author} {\bibfnamefont {I.~A.}\ \bibnamefont
  {Mart{\'{i}}nez}}, \bibinfo {author} {\bibfnamefont {J.~M.}\ \bibnamefont
  {Parrondo}},\ and\ \bibinfo {author} {\bibfnamefont {D.}~\bibnamefont
  {Petrov}},\ }\bibfield  {title} {\bibinfo {title} {{Universal features in the
  energetics of symmetry breaking}},\ }\href
  {https://doi.org/10.1038/nphys2940} {\bibfield  {journal} {\bibinfo
  {journal} {Nature Physics}\ }\textbf {\bibinfo {volume} {10}},\ \bibinfo
  {pages} {457} (\bibinfo {year} {2014})},\ \Eprint
  {https://arxiv.org/abs/1310.5518} {arXiv:1310.5518} \BibitemShut {NoStop}%
\bibitem [{\citenamefont {Koski}\ \emph
  {et~al.}(2014{\natexlab{a}})\citenamefont {Koski}, \citenamefont {Maisi},
  \citenamefont {Sagawa},\ and\ \citenamefont {Pekola}}]{Koski2014}%
  \BibitemOpen
  \bibfield  {author} {\bibinfo {author} {\bibfnamefont {J.~V.}\ \bibnamefont
  {Koski}}, \bibinfo {author} {\bibfnamefont {V.~F.}\ \bibnamefont {Maisi}},
  \bibinfo {author} {\bibfnamefont {T.}~\bibnamefont {Sagawa}},\ and\ \bibinfo
  {author} {\bibfnamefont {J.~P.}\ \bibnamefont {Pekola}},\ }\bibfield  {title}
  {\bibinfo {title} {{Experimental observation of the role of mutual
  information in the nonequilibrium dynamics of a Maxwell demon}},\ }\href
  {https://doi.org/10.1103/PhysRevLett.113.030601} {\bibfield  {journal}
  {\bibinfo  {journal} {Physical Review Letters}\ }\textbf {\bibinfo {volume}
  {113}},\ \bibinfo {pages} {030601} (\bibinfo {year}
  {2014}{\natexlab{a}})}\BibitemShut {NoStop}%
\bibitem [{\citenamefont {Koski}\ \emph
  {et~al.}(2014{\natexlab{b}})\citenamefont {Koski}, \citenamefont {Maisi},
  \citenamefont {Pekola},\ and\ \citenamefont {Averin}}]{Koski2014-PNAS}%
  \BibitemOpen
  \bibfield  {author} {\bibinfo {author} {\bibfnamefont {J.~V.}\ \bibnamefont
  {Koski}}, \bibinfo {author} {\bibfnamefont {V.~F.}\ \bibnamefont {Maisi}},
  \bibinfo {author} {\bibfnamefont {J.~P.}\ \bibnamefont {Pekola}},\ and\
  \bibinfo {author} {\bibfnamefont {D.~V.}\ \bibnamefont {Averin}},\ }\bibfield
   {title} {\bibinfo {title} {{Experimental realization of a Szilard engine
  with a single electron}},\ }\href {https://doi.org/10.1073/pnas.1406966111}
  {\bibfield  {journal} {\bibinfo  {journal} {Proceedings of the National
  Academy of Sciences of the United States of America}\ }\textbf {\bibinfo
  {volume} {111}},\ \bibinfo {pages} {13786} (\bibinfo {year}
  {2014}{\natexlab{b}})},\ \Eprint {https://arxiv.org/abs/1402.5907}
  {arXiv:1402.5907} \BibitemShut {NoStop}%
\bibitem [{\citenamefont {Vidrighin}\ \emph {et~al.}(2016)\citenamefont
  {Vidrighin}, \citenamefont {Dahlsten}, \citenamefont {Barbieri},
  \citenamefont {Kim}, \citenamefont {Vedral},\ and\ \citenamefont
  {Walmsley}}]{Vidrighin2016}%
  \BibitemOpen
  \bibfield  {author} {\bibinfo {author} {\bibfnamefont {M.~D.}\ \bibnamefont
  {Vidrighin}}, \bibinfo {author} {\bibfnamefont {O.}~\bibnamefont {Dahlsten}},
  \bibinfo {author} {\bibfnamefont {M.}~\bibnamefont {Barbieri}}, \bibinfo
  {author} {\bibfnamefont {M.~S.}\ \bibnamefont {Kim}}, \bibinfo {author}
  {\bibfnamefont {V.}~\bibnamefont {Vedral}},\ and\ \bibinfo {author}
  {\bibfnamefont {I.~A.}\ \bibnamefont {Walmsley}},\ }\bibfield  {title}
  {\bibinfo {title} {{Photonic Maxwell's Demon}},\ }\href
  {https://doi.org/10.1103/PhysRevLett.116.050401} {\bibfield  {journal}
  {\bibinfo  {journal} {Physical Review Letters}\ }\textbf {\bibinfo {volume}
  {116}},\ \bibinfo {pages} {1} (\bibinfo {year} {2016})},\ \Eprint
  {https://arxiv.org/abs/1510.02164} {arXiv:1510.02164} \BibitemShut {NoStop}%
\bibitem [{\citenamefont {Camati}\ \emph {et~al.}(2016)\citenamefont {Camati},
  \citenamefont {Peterson}, \citenamefont {Batalh{\~{a}}o}, \citenamefont
  {Micadei}, \citenamefont {Souza}, \citenamefont {Sarthour}, \citenamefont
  {Oliveira},\ and\ \citenamefont {Serra}}]{Camati2016}%
  \BibitemOpen
  \bibfield  {author} {\bibinfo {author} {\bibfnamefont {P.~A.}\ \bibnamefont
  {Camati}}, \bibinfo {author} {\bibfnamefont {J.~P.}\ \bibnamefont
  {Peterson}}, \bibinfo {author} {\bibfnamefont {T.~B.}\ \bibnamefont
  {Batalh{\~{a}}o}}, \bibinfo {author} {\bibfnamefont {K.}~\bibnamefont
  {Micadei}}, \bibinfo {author} {\bibfnamefont {A.~M.}\ \bibnamefont {Souza}},
  \bibinfo {author} {\bibfnamefont {R.~S.}\ \bibnamefont {Sarthour}}, \bibinfo
  {author} {\bibfnamefont {I.~S.}\ \bibnamefont {Oliveira}},\ and\ \bibinfo
  {author} {\bibfnamefont {R.~M.}\ \bibnamefont {Serra}},\ }\bibfield  {title}
  {\bibinfo {title} {{Experimental Rectification of Entropy Production by
  Maxwell's Demon in a Quantum System}},\ }\href
  {https://doi.org/10.1103/PhysRevLett.117.240502} {\bibfield  {journal}
  {\bibinfo  {journal} {Physical Review Letters}\ }\textbf {\bibinfo {volume}
  {117}},\ \bibinfo {pages} {240502} (\bibinfo {year} {2016})},\ \Eprint
  {https://arxiv.org/abs/1605.08821} {arXiv:1605.08821} \BibitemShut {NoStop}%
\bibitem [{\citenamefont {Ciampini}\ \emph {et~al.}(2017)\citenamefont
  {Ciampini}, \citenamefont {Mancino}, \citenamefont {Orieux}, \citenamefont
  {Vigliar}, \citenamefont {Mataloni}, \citenamefont {Paternostro},\ and\
  \citenamefont {Barbieri}}]{Ciampini2017}%
  \BibitemOpen
  \bibfield  {author} {\bibinfo {author} {\bibfnamefont {M.~A.}\ \bibnamefont
  {Ciampini}}, \bibinfo {author} {\bibfnamefont {L.}~\bibnamefont {Mancino}},
  \bibinfo {author} {\bibfnamefont {A.}~\bibnamefont {Orieux}}, \bibinfo
  {author} {\bibfnamefont {C.}~\bibnamefont {Vigliar}}, \bibinfo {author}
  {\bibfnamefont {P.}~\bibnamefont {Mataloni}}, \bibinfo {author}
  {\bibfnamefont {M.}~\bibnamefont {Paternostro}},\ and\ \bibinfo {author}
  {\bibfnamefont {M.}~\bibnamefont {Barbieri}},\ }\bibfield  {title} {\bibinfo
  {title} {{Experimental extractable work-based multipartite separability
  criteria}},\ }\href {https://doi.org/10.1038/s41534-017-0011-9} {\bibfield
  {journal} {\bibinfo  {journal} {npj Quantum Information}\ }\textbf {\bibinfo
  {volume} {3}},\ \bibinfo {pages} {10} (\bibinfo {year} {2017})}\BibitemShut
  {NoStop}%
\bibitem [{\citenamefont {Cottet}\ \emph {et~al.}(2017)\citenamefont {Cottet},
  \citenamefont {Jezouin}, \citenamefont {Bretheau}, \citenamefont
  {Campagne-Ibarcq}, \citenamefont {Ficheux}, \citenamefont {Anders},
  \citenamefont {Auff{\`{e}}ves}, \citenamefont {Azouit}, \citenamefont
  {Rouchon},\ and\ \citenamefont {Huard}}]{Cottet2017}%
  \BibitemOpen
  \bibfield  {author} {\bibinfo {author} {\bibfnamefont {N.}~\bibnamefont
  {Cottet}}, \bibinfo {author} {\bibfnamefont {S.}~\bibnamefont {Jezouin}},
  \bibinfo {author} {\bibfnamefont {L.}~\bibnamefont {Bretheau}}, \bibinfo
  {author} {\bibfnamefont {P.}~\bibnamefont {Campagne-Ibarcq}}, \bibinfo
  {author} {\bibfnamefont {Q.}~\bibnamefont {Ficheux}}, \bibinfo {author}
  {\bibfnamefont {J.}~\bibnamefont {Anders}}, \bibinfo {author} {\bibfnamefont
  {A.}~\bibnamefont {Auff{\`{e}}ves}}, \bibinfo {author} {\bibfnamefont
  {R.}~\bibnamefont {Azouit}}, \bibinfo {author} {\bibfnamefont
  {P.}~\bibnamefont {Rouchon}},\ and\ \bibinfo {author} {\bibfnamefont
  {B.}~\bibnamefont {Huard}},\ }\bibfield  {title} {\bibinfo {title}
  {{Observing a quantum Maxwell demon at work}},\ }\href
  {https://doi.org/10.1073/pnas.1704827114} {\bibfield  {journal} {\bibinfo
  {journal} {Proceedings of the National Academy of Sciences of the United
  States of America}\ }\textbf {\bibinfo {volume} {114}},\ \bibinfo {pages}
  {7561} (\bibinfo {year} {2017})},\ \Eprint {https://arxiv.org/abs/1702.05161}
  {arXiv:1702.05161} \BibitemShut {NoStop}%
\bibitem [{\citenamefont {Masuyama}\ \emph {et~al.}(2018)\citenamefont
  {Masuyama}, \citenamefont {Funo}, \citenamefont {Murashita}, \citenamefont
  {Noguchi}, \citenamefont {Kono}, \citenamefont {Tabuchi}, \citenamefont
  {Yamazaki}, \citenamefont {Ueda},\ and\ \citenamefont
  {Nakamura}}]{Masuyama2018}%
  \BibitemOpen
  \bibfield  {author} {\bibinfo {author} {\bibfnamefont {Y.}~\bibnamefont
  {Masuyama}}, \bibinfo {author} {\bibfnamefont {K.}~\bibnamefont {Funo}},
  \bibinfo {author} {\bibfnamefont {Y.}~\bibnamefont {Murashita}}, \bibinfo
  {author} {\bibfnamefont {A.}~\bibnamefont {Noguchi}}, \bibinfo {author}
  {\bibfnamefont {S.}~\bibnamefont {Kono}}, \bibinfo {author} {\bibfnamefont
  {Y.}~\bibnamefont {Tabuchi}}, \bibinfo {author} {\bibfnamefont
  {R.}~\bibnamefont {Yamazaki}}, \bibinfo {author} {\bibfnamefont
  {M.}~\bibnamefont {Ueda}},\ and\ \bibinfo {author} {\bibfnamefont
  {Y.}~\bibnamefont {Nakamura}},\ }\bibfield  {title} {\bibinfo {title}
  {{Information-to-work conversion by Maxwell's demon in a superconducting
  circuit quantum electrodynamical system}},\ }\href
  {https://doi.org/10.1038/s41467-018-03686-y} {\bibfield  {journal} {\bibinfo
  {journal} {Nature Communications}\ }\textbf {\bibinfo {volume} {9}},\
  \bibinfo {pages} {1291} (\bibinfo {year} {2018})},\ \Eprint
  {https://arxiv.org/abs/1709.00548} {arXiv:1709.00548} \BibitemShut {NoStop}%
\bibitem [{\citenamefont {Annby-Andersson}\ \emph {et~al.}(2020)\citenamefont
  {Annby-Andersson}, \citenamefont {Samuelsson}, \citenamefont {Maisi},\ and\
  \citenamefont {Potts}}]{Annby-Andersson2020}%
  \BibitemOpen
  \bibfield  {author} {\bibinfo {author} {\bibfnamefont {B.}~\bibnamefont
  {Annby-Andersson}}, \bibinfo {author} {\bibfnamefont {P.}~\bibnamefont
  {Samuelsson}}, \bibinfo {author} {\bibfnamefont {V.~F.}\ \bibnamefont
  {Maisi}},\ and\ \bibinfo {author} {\bibfnamefont {P.~P.}\ \bibnamefont
  {Potts}},\ }\bibfield  {title} {\bibinfo {title} {{Maxwell's demon in a
  double quantum dot with continuous charge detection}},\ }\href
  {https://doi.org/10.1103/PhysRevB.101.165404} {\bibfield  {journal} {\bibinfo
   {journal} {Physical Review B}\ }\textbf {\bibinfo {volume} {101}},\ \bibinfo
  {pages} {165404} (\bibinfo {year} {2020})},\ \Eprint
  {https://arxiv.org/abs/1912.09188} {arXiv:1912.09188} \BibitemShut {NoStop}%
\bibitem [{\citenamefont {Paule}\ \emph {et~al.}(2020)\citenamefont {Paule},
  \citenamefont {Subero}, \citenamefont {Maillet}, \citenamefont {Fazio},
  \citenamefont {Pekola},\ and\ \citenamefont {Rold{\'{a}}n}}]{Manzano2020}%
  \BibitemOpen
  \bibfield  {author} {\bibinfo {author} {\bibfnamefont {G.~M.}\ \bibnamefont
  {Paule}}, \bibinfo {author} {\bibfnamefont {D.}~\bibnamefont {Subero}},
  \bibinfo {author} {\bibfnamefont {O.}~\bibnamefont {Maillet}}, \bibinfo
  {author} {\bibfnamefont {R.}~\bibnamefont {Fazio}}, \bibinfo {author}
  {\bibfnamefont {J.~P.}\ \bibnamefont {Pekola}},\ and\ \bibinfo {author}
  {\bibfnamefont {{\'{E}}.}~\bibnamefont {Rold{\'{a}}n}},\ }\href
  {http://arxiv.org/abs/2008.01630} {\bibinfo {title} {{Thermodynamics of
  gambling demons}}} (\bibinfo {year} {2020}),\ \Eprint
  {https://arxiv.org/abs/2008.01630} {arXiv:2008.01630} \BibitemShut {NoStop}%
\bibitem [{\citenamefont {Naghiloo}\ \emph {et~al.}(2018)\citenamefont
  {Naghiloo}, \citenamefont {Alonso}, \citenamefont {Romito}, \citenamefont
  {Lutz},\ and\ \citenamefont {Murch}}]{Naghiloo2018}%
  \BibitemOpen
  \bibfield  {author} {\bibinfo {author} {\bibfnamefont {M.}~\bibnamefont
  {Naghiloo}}, \bibinfo {author} {\bibfnamefont {J.~J.}\ \bibnamefont
  {Alonso}}, \bibinfo {author} {\bibfnamefont {A.}~\bibnamefont {Romito}},
  \bibinfo {author} {\bibfnamefont {E.}~\bibnamefont {Lutz}},\ and\ \bibinfo
  {author} {\bibfnamefont {K.~W.}\ \bibnamefont {Murch}},\ }\bibfield  {title}
  {\bibinfo {title} {{Information Gain and Loss for a Quantum Maxwell's
  Demon}},\ }\href {https://doi.org/10.1103/PhysRevLett.121.030604} {\bibfield
  {journal} {\bibinfo  {journal} {Physical Review Letters}\ }\textbf {\bibinfo
  {volume} {121}},\ \bibinfo {pages} {030604} (\bibinfo {year} {2018})},\
  \Eprint {https://arxiv.org/abs/1802.07205} {arXiv:1802.07205} \BibitemShut
  {NoStop}%
\bibitem [{\citenamefont {Najera-Santos}\ \emph {et~al.}(2020)\citenamefont
  {Najera-Santos}, \citenamefont {Camati}, \citenamefont {M{\'{e}}tillon},
  \citenamefont {Brune}, \citenamefont {Raimond}, \citenamefont
  {Auff{\`{e}}ves},\ and\ \citenamefont {Dotsenko}}]{Najera-Santos2020}%
  \BibitemOpen
  \bibfield  {author} {\bibinfo {author} {\bibfnamefont {B.-L.}\ \bibnamefont
  {Najera-Santos}}, \bibinfo {author} {\bibfnamefont {P.~A.}\ \bibnamefont
  {Camati}}, \bibinfo {author} {\bibfnamefont {V.}~\bibnamefont
  {M{\'{e}}tillon}}, \bibinfo {author} {\bibfnamefont {M.}~\bibnamefont
  {Brune}}, \bibinfo {author} {\bibfnamefont {J.-M.}\ \bibnamefont {Raimond}},
  \bibinfo {author} {\bibfnamefont {A.}~\bibnamefont {Auff{\`{e}}ves}},\ and\
  \bibinfo {author} {\bibfnamefont {I.}~\bibnamefont {Dotsenko}},\ }\bibfield
  {title} {\bibinfo {title} {{Autonomous Maxwell's demon in a cavity QED
  system}},\ }\href {https://doi.org/10.1103/PhysRevResearch.2.032025}
  {\bibfield  {journal} {\bibinfo  {journal} {Physical Review Research}\
  }\textbf {\bibinfo {volume} {2}},\ \bibinfo {pages} {032025} (\bibinfo {year}
  {2020})},\ \Eprint {https://arxiv.org/abs/2001.07445} {arXiv:2001.07445}
  \BibitemShut {NoStop}%
\bibitem [{\citenamefont {S{\'{a}}nchez}\ \emph {et~al.}(2019)\citenamefont
  {S{\'{a}}nchez}, \citenamefont {Samuelsson},\ and\ \citenamefont
  {Potts}}]{Sanchez2019}%
  \BibitemOpen
  \bibfield  {author} {\bibinfo {author} {\bibfnamefont {R.}~\bibnamefont
  {S{\'{a}}nchez}}, \bibinfo {author} {\bibfnamefont {P.}~\bibnamefont
  {Samuelsson}},\ and\ \bibinfo {author} {\bibfnamefont {P.~P.}\ \bibnamefont
  {Potts}},\ }\bibfield  {title} {\bibinfo {title} {{Autonomous conversion of
  information to work in quantum dots}},\ }\href
  {https://doi.org/10.1103/PhysRevResearch.1.033066} {\bibfield  {journal}
  {\bibinfo  {journal} {Physical Review Research}\ }\textbf {\bibinfo {volume}
  {1}},\ \bibinfo {pages} {033066} (\bibinfo {year} {2019})},\ \Eprint
  {https://arxiv.org/abs/1907.02866} {arXiv:1907.02866} \BibitemShut {NoStop}%
\bibitem [{\citenamefont {Kumar}\ \emph {et~al.}(2018)\citenamefont {Kumar},
  \citenamefont {Wu}, \citenamefont {Giraldo},\ and\ \citenamefont
  {Weiss}}]{Kumar2018}%
  \BibitemOpen
  \bibfield  {author} {\bibinfo {author} {\bibfnamefont {A.}~\bibnamefont
  {Kumar}}, \bibinfo {author} {\bibfnamefont {T.~Y.}\ \bibnamefont {Wu}},
  \bibinfo {author} {\bibfnamefont {F.}~\bibnamefont {Giraldo}},\ and\ \bibinfo
  {author} {\bibfnamefont {D.~S.}\ \bibnamefont {Weiss}},\ }\bibfield  {title}
  {\bibinfo {title} {{Sorting ultracold atoms in a three-dimensional optical
  lattice in a realization of Maxwell's demon}},\ }\href
  {https://doi.org/10.1038/s41586-018-0458-7} {\bibfield  {journal} {\bibinfo
  {journal} {Nature}\ }\textbf {\bibinfo {volume} {561}},\ \bibinfo {pages}
  {83} (\bibinfo {year} {2018})}\BibitemShut {NoStop}%
\bibitem [{\citenamefont {Chow}\ \emph {et~al.}(2011)\citenamefont {Chow},
  \citenamefont {C{\'{o}}rcoles}, \citenamefont {Gambetta}, \citenamefont
  {Rigetti}, \citenamefont {Johnson}, \citenamefont {Smolin}, \citenamefont
  {Rozen}, \citenamefont {Keefe}, \citenamefont {Rothwell}, \citenamefont
  {Ketchen},\ and\ \citenamefont {Steffen}}]{Chow2011}%
  \BibitemOpen
  \bibfield  {author} {\bibinfo {author} {\bibfnamefont {J.~M.}\ \bibnamefont
  {Chow}}, \bibinfo {author} {\bibfnamefont {A.~D.}\ \bibnamefont
  {C{\'{o}}rcoles}}, \bibinfo {author} {\bibfnamefont {J.~M.}\ \bibnamefont
  {Gambetta}}, \bibinfo {author} {\bibfnamefont {C.}~\bibnamefont {Rigetti}},
  \bibinfo {author} {\bibfnamefont {B.~R.}\ \bibnamefont {Johnson}}, \bibinfo
  {author} {\bibfnamefont {J.~A.}\ \bibnamefont {Smolin}}, \bibinfo {author}
  {\bibfnamefont {J.~R.}\ \bibnamefont {Rozen}}, \bibinfo {author}
  {\bibfnamefont {G.~A.}\ \bibnamefont {Keefe}}, \bibinfo {author}
  {\bibfnamefont {M.~B.}\ \bibnamefont {Rothwell}}, \bibinfo {author}
  {\bibfnamefont {M.~B.}\ \bibnamefont {Ketchen}},\ and\ \bibinfo {author}
  {\bibfnamefont {M.}~\bibnamefont {Steffen}},\ }\bibfield  {title} {\bibinfo
  {title} {{Simple all-microwave entangling gate for fixed-frequency
  superconducting qubits}},\ }\href
  {https://doi.org/10.1103/PhysRevLett.107.080502} {\bibfield  {journal}
  {\bibinfo  {journal} {Physical Review Letters}\ }\textbf {\bibinfo {volume}
  {107}},\ \bibinfo {pages} {080502} (\bibinfo {year} {2011})},\ \Eprint
  {https://arxiv.org/abs/1106.0553} {arXiv:1106.0553} \BibitemShut {NoStop}%
\bibitem [{\citenamefont {{De Lange}}\ \emph {et~al.}(2010)\citenamefont {{De
  Lange}}, \citenamefont {Wang}, \citenamefont {Rist{\`{e}}}, \citenamefont
  {Dobrovitski},\ and\ \citenamefont {Hanson}}]{Martinac2010}%
  \BibitemOpen
  \bibfield  {author} {\bibinfo {author} {\bibfnamefont {G.}~\bibnamefont {{De
  Lange}}}, \bibinfo {author} {\bibfnamefont {Z.~H.}\ \bibnamefont {Wang}},
  \bibinfo {author} {\bibfnamefont {D.}~\bibnamefont {Rist{\`{e}}}}, \bibinfo
  {author} {\bibfnamefont {V.~V.}\ \bibnamefont {Dobrovitski}},\ and\ \bibinfo
  {author} {\bibfnamefont {R.}~\bibnamefont {Hanson}},\ }\bibfield  {title}
  {\bibinfo {title} {{Universal dynamical decoupling of a single solid-state
  spin from a spin bath}},\ }\href {https://doi.org/10.1126/science.1192739}
  {\bibfield  {journal} {\bibinfo  {journal} {Science}\ }\textbf {\bibinfo
  {volume} {330}},\ \bibinfo {pages} {60} (\bibinfo {year} {2010})},\ \Eprint
  {https://arxiv.org/abs/1008.2119} {arXiv:1008.2119} \BibitemShut {NoStop}%
\bibitem [{\citenamefont {Rosenblum}\ \emph {et~al.}(2018)\citenamefont
  {Rosenblum}, \citenamefont {Gao}, \citenamefont {Reinhold}, \citenamefont
  {Wang}, \citenamefont {Axline}, \citenamefont {Frunzio}, \citenamefont
  {Girvin}, \citenamefont {Jiang}, \citenamefont {Mirrahimi}, \citenamefont
  {Devoret},\ and\ \citenamefont {Schoelkopf}}]{Rosenblum2018}%
  \BibitemOpen
  \bibfield  {author} {\bibinfo {author} {\bibfnamefont {S.}~\bibnamefont
  {Rosenblum}}, \bibinfo {author} {\bibfnamefont {Y.~Y.}\ \bibnamefont {Gao}},
  \bibinfo {author} {\bibfnamefont {P.}~\bibnamefont {Reinhold}}, \bibinfo
  {author} {\bibfnamefont {C.}~\bibnamefont {Wang}}, \bibinfo {author}
  {\bibfnamefont {C.~J.}\ \bibnamefont {Axline}}, \bibinfo {author}
  {\bibfnamefont {L.}~\bibnamefont {Frunzio}}, \bibinfo {author} {\bibfnamefont
  {S.~M.}\ \bibnamefont {Girvin}}, \bibinfo {author} {\bibfnamefont
  {L.}~\bibnamefont {Jiang}}, \bibinfo {author} {\bibfnamefont
  {M.}~\bibnamefont {Mirrahimi}}, \bibinfo {author} {\bibfnamefont {M.~H.}\
  \bibnamefont {Devoret}},\ and\ \bibinfo {author} {\bibfnamefont {R.~J.}\
  \bibnamefont {Schoelkopf}},\ }\bibfield  {title} {\bibinfo {title} {{A CNOT
  gate between multiphoton qubits encoded in two cavities}},\ }\href
  {https://doi.org/10.1038/s41467-018-03059-5} {\bibfield  {journal} {\bibinfo
  {journal} {Nature Communications}\ }\textbf {\bibinfo {volume} {9}},\
  \bibinfo {pages} {652} (\bibinfo {year} {2018})},\ \Eprint
  {https://arxiv.org/abs/1709.05425} {arXiv:1709.05425} \BibitemShut {NoStop}%
\bibitem [{\citenamefont {Riebe}\ \emph {et~al.}(2006)\citenamefont {Riebe},
  \citenamefont {Kim}, \citenamefont {Schindler}, \citenamefont {Monz},
  \citenamefont {Schmidt}, \citenamefont {K{\"{o}}rber}, \citenamefont
  {H{\"{a}}nsel}, \citenamefont {H{\"{a}}ffner}, \citenamefont {Roos},\ and\
  \citenamefont {Blatt}}]{Riebe2006}%
  \BibitemOpen
  \bibfield  {author} {\bibinfo {author} {\bibfnamefont {M.}~\bibnamefont
  {Riebe}}, \bibinfo {author} {\bibfnamefont {K.}~\bibnamefont {Kim}}, \bibinfo
  {author} {\bibfnamefont {P.}~\bibnamefont {Schindler}}, \bibinfo {author}
  {\bibfnamefont {T.}~\bibnamefont {Monz}}, \bibinfo {author} {\bibfnamefont
  {P.~O.}\ \bibnamefont {Schmidt}}, \bibinfo {author} {\bibfnamefont {T.~K.}\
  \bibnamefont {K{\"{o}}rber}}, \bibinfo {author} {\bibfnamefont
  {W.}~\bibnamefont {H{\"{a}}nsel}}, \bibinfo {author} {\bibfnamefont
  {H.}~\bibnamefont {H{\"{a}}ffner}}, \bibinfo {author} {\bibfnamefont {C.~F.}\
  \bibnamefont {Roos}},\ and\ \bibinfo {author} {\bibfnamefont
  {R.}~\bibnamefont {Blatt}},\ }\bibfield  {title} {\bibinfo {title} {{Process
  tomography of ion trap quantum gates}},\ }\href
  {https://doi.org/10.1103/PhysRevLett.97.220407} {\bibfield  {journal}
  {\bibinfo  {journal} {Physical Review Letters}\ }\textbf {\bibinfo {volume}
  {97}},\ \bibinfo {pages} {220407} (\bibinfo {year} {2006})},\ \Eprint
  {https://arxiv.org/abs/quant-ph/0609228} {arXiv:quant-ph/0609228}
  \BibitemShut {NoStop}%
\bibitem [{\citenamefont {Yamamoto}\ \emph {et~al.}(2010)\citenamefont
  {Yamamoto}, \citenamefont {Neeley}, \citenamefont {Lucero}, \citenamefont
  {Bialczak}, \citenamefont {Kelly}, \citenamefont {Lenander}, \citenamefont
  {Mariantoni}, \citenamefont {O'Connell}, \citenamefont {Sank}, \citenamefont
  {Wang}, \citenamefont {Weides}, \citenamefont {Wenner}, \citenamefont {Yin},
  \citenamefont {Cleland},\ and\ \citenamefont {Martinis}}]{Yamamoto2010}%
  \BibitemOpen
  \bibfield  {author} {\bibinfo {author} {\bibfnamefont {T.}~\bibnamefont
  {Yamamoto}}, \bibinfo {author} {\bibfnamefont {M.}~\bibnamefont {Neeley}},
  \bibinfo {author} {\bibfnamefont {E.}~\bibnamefont {Lucero}}, \bibinfo
  {author} {\bibfnamefont {R.~C.}\ \bibnamefont {Bialczak}}, \bibinfo {author}
  {\bibfnamefont {J.}~\bibnamefont {Kelly}}, \bibinfo {author} {\bibfnamefont
  {M.}~\bibnamefont {Lenander}}, \bibinfo {author} {\bibfnamefont
  {M.}~\bibnamefont {Mariantoni}}, \bibinfo {author} {\bibfnamefont {A.~D.}\
  \bibnamefont {O'Connell}}, \bibinfo {author} {\bibfnamefont {D.}~\bibnamefont
  {Sank}}, \bibinfo {author} {\bibfnamefont {H.}~\bibnamefont {Wang}}, \bibinfo
  {author} {\bibfnamefont {M.}~\bibnamefont {Weides}}, \bibinfo {author}
  {\bibfnamefont {J.}~\bibnamefont {Wenner}}, \bibinfo {author} {\bibfnamefont
  {Y.}~\bibnamefont {Yin}}, \bibinfo {author} {\bibfnamefont {A.~N.}\
  \bibnamefont {Cleland}},\ and\ \bibinfo {author} {\bibfnamefont {J.~M.}\
  \bibnamefont {Martinis}},\ }\bibfield  {title} {\bibinfo {title} {{Quantum
  process tomography of two-qubit controlled-Z and controlled-NOT gates using
  superconducting phase qubits}},\ }\href
  {https://doi.org/10.1103/PhysRevB.82.184515} {\bibfield  {journal} {\bibinfo
  {journal} {Physical Review B - Condensed Matter and Materials Physics}\
  }\textbf {\bibinfo {volume} {82}},\ \bibinfo {pages} {184515} (\bibinfo
  {year} {2010})},\ \Eprint {https://arxiv.org/abs/1006.5084} {arXiv:1006.5084}
  \BibitemShut {NoStop}%
\bibitem [{\citenamefont {Childs}\ \emph {et~al.}(2001)\citenamefont {Childs},
  \citenamefont {Chuang},\ and\ \citenamefont {Leung}}]{Childs2001}%
  \BibitemOpen
  \bibfield  {author} {\bibinfo {author} {\bibfnamefont {A.~M.}\ \bibnamefont
  {Childs}}, \bibinfo {author} {\bibfnamefont {I.~L.}\ \bibnamefont {Chuang}},\
  and\ \bibinfo {author} {\bibfnamefont {D.~W.}\ \bibnamefont {Leung}},\
  }\bibfield  {title} {\bibinfo {title} {{Realization of quantum process
  tomography in NMR}},\ }\href {https://doi.org/10.1103/PhysRevA.64.012314}
  {\bibfield  {journal} {\bibinfo  {journal} {Physical Review A. Atomic,
  Molecular, and Optical Physics}\ }\textbf {\bibinfo {volume} {64}},\ \bibinfo
  {pages} {123141} (\bibinfo {year} {2001})},\ \Eprint
  {https://arxiv.org/abs/quant-ph/0012032} {arXiv:quant-ph/0012032}
  \BibitemShut {NoStop}%
\bibitem [{\citenamefont {Kupchak}\ \emph {et~al.}(2015)\citenamefont
  {Kupchak}, \citenamefont {Rind}, \citenamefont {Jordaan},\ and\ \citenamefont
  {Figueroa}}]{Kupchak2015}%
  \BibitemOpen
  \bibfield  {author} {\bibinfo {author} {\bibfnamefont {C.}~\bibnamefont
  {Kupchak}}, \bibinfo {author} {\bibfnamefont {S.}~\bibnamefont {Rind}},
  \bibinfo {author} {\bibfnamefont {B.}~\bibnamefont {Jordaan}},\ and\ \bibinfo
  {author} {\bibfnamefont {E.}~\bibnamefont {Figueroa}},\ }\bibfield  {title}
  {\bibinfo {title} {{Quantum Process Tomography of an Optically-Controlled
  Kerr Non-linearity}},\ }\href {https://doi.org/10.1038/srep16581} {\bibfield
  {journal} {\bibinfo  {journal} {Scientific Reports}\ }\textbf {\bibinfo
  {volume} {5}},\ \bibinfo {pages} {16581} (\bibinfo {year} {2015})},\ \Eprint
  {https://arxiv.org/abs/1505.03918} {arXiv:1505.03918} \BibitemShut {NoStop}%
\bibitem [{\citenamefont {Kim}\ \emph {et~al.}(2018)\citenamefont {Kim},
  \citenamefont {Kim}, \citenamefont {Lee}, \citenamefont {Han}, \citenamefont
  {Moon}, \citenamefont {Kim},\ and\ \citenamefont {Cho}}]{Kim2018}%
  \BibitemOpen
  \bibfield  {author} {\bibinfo {author} {\bibfnamefont {Y.}~\bibnamefont
  {Kim}}, \bibinfo {author} {\bibfnamefont {Y.~S.}\ \bibnamefont {Kim}},
  \bibinfo {author} {\bibfnamefont {S.~Y.}\ \bibnamefont {Lee}}, \bibinfo
  {author} {\bibfnamefont {S.~W.}\ \bibnamefont {Han}}, \bibinfo {author}
  {\bibfnamefont {S.}~\bibnamefont {Moon}}, \bibinfo {author} {\bibfnamefont
  {Y.~H.}\ \bibnamefont {Kim}},\ and\ \bibinfo {author} {\bibfnamefont {Y.~W.}\
  \bibnamefont {Cho}},\ }\bibfield  {title} {\bibinfo {title} {{Direct quantum
  process tomography via measuring sequential weak values of incompatible
  observables}},\ }\href {https://doi.org/10.1038/s41467-017-02511-2}
  {\bibfield  {journal} {\bibinfo  {journal} {Nature Communications}\ }\textbf
  {\bibinfo {volume} {9}},\ \bibinfo {pages} {192} (\bibinfo {year}
  {2018})}\BibitemShut {NoStop}%
\bibitem [{\citenamefont {Altepeter}\ \emph {et~al.}(2003)\citenamefont
  {Altepeter}, \citenamefont {Branning}, \citenamefont {Jeffrey}, \citenamefont
  {Wei}, \citenamefont {Kwiat}, \citenamefont {Thew}, \citenamefont {O'Brien},
  \citenamefont {Nielsen},\ and\ \citenamefont {White}}]{Altepeter2003}%
  \BibitemOpen
  \bibfield  {author} {\bibinfo {author} {\bibfnamefont {J.~B.}\ \bibnamefont
  {Altepeter}}, \bibinfo {author} {\bibfnamefont {D.}~\bibnamefont {Branning}},
  \bibinfo {author} {\bibfnamefont {E.}~\bibnamefont {Jeffrey}}, \bibinfo
  {author} {\bibfnamefont {T.~C.}\ \bibnamefont {Wei}}, \bibinfo {author}
  {\bibfnamefont {P.~G.}\ \bibnamefont {Kwiat}}, \bibinfo {author}
  {\bibfnamefont {R.~T.}\ \bibnamefont {Thew}}, \bibinfo {author}
  {\bibfnamefont {J.~L.}\ \bibnamefont {O'Brien}}, \bibinfo {author}
  {\bibfnamefont {M.~A.}\ \bibnamefont {Nielsen}},\ and\ \bibinfo {author}
  {\bibfnamefont {A.~G.}\ \bibnamefont {White}},\ }\bibfield  {title} {\bibinfo
  {title} {{Ancilla-Assisted Quantum Process Tomography}},\ }\href
  {https://doi.org/10.1103/PhysRevLett.90.193601} {\bibfield  {journal}
  {\bibinfo  {journal} {Physical Review Letters}\ }\textbf {\bibinfo {volume}
  {90}},\ \bibinfo {pages} {4} (\bibinfo {year} {2003})},\ \Eprint
  {https://arxiv.org/abs/quant-ph/0303038} {arXiv:quant-ph/0303038}
  \BibitemShut {NoStop}%
\bibitem [{\citenamefont {Horowitz}\ and\ \citenamefont
  {Parrondo}(2011)}]{Horowitz2011}%
  \BibitemOpen
  \bibfield  {author} {\bibinfo {author} {\bibfnamefont {J.~M.}\ \bibnamefont
  {Horowitz}}\ and\ \bibinfo {author} {\bibfnamefont {J.~M.}\ \bibnamefont
  {Parrondo}},\ }\bibfield  {title} {\bibinfo {title} {{Designing optimal
  discrete-feedback thermodynamic engines}},\ }\href
  {https://doi.org/10.1088/1367-2630/13/12/123019} {\bibfield  {journal}
  {\bibinfo  {journal} {New Journal of Physics}\ }\textbf {\bibinfo {volume}
  {13}},\ \bibinfo {pages} {123019} (\bibinfo {year} {2011})},\ \Eprint
  {https://arxiv.org/abs/1110.6808} {arXiv:1110.6808} \BibitemShut {NoStop}%
\bibitem [{\citenamefont {Manzano}\ \emph {et~al.}(2018)\citenamefont
  {Manzano}, \citenamefont {Plastina},\ and\ \citenamefont
  {Zambrini}}]{Manzano2018}%
  \BibitemOpen
  \bibfield  {author} {\bibinfo {author} {\bibfnamefont {G.}~\bibnamefont
  {Manzano}}, \bibinfo {author} {\bibfnamefont {F.}~\bibnamefont {Plastina}},\
  and\ \bibinfo {author} {\bibfnamefont {R.}~\bibnamefont {Zambrini}},\
  }\bibfield  {title} {\bibinfo {title} {{Optimal Work Extraction and
  Thermodynamics of Quantum Measurements and Correlations}},\ }\href
  {https://doi.org/10.1103/PhysRevLett.121.120602} {\bibfield  {journal}
  {\bibinfo  {journal} {Physical Review Letters}\ }\textbf {\bibinfo {volume}
  {121}},\ \bibinfo {pages} {120602} (\bibinfo {year} {2018})},\ \Eprint
  {https://arxiv.org/abs/1805.08184} {arXiv:1805.08184} \BibitemShut {NoStop}%
\bibitem [{\citenamefont {Paneru}\ \emph {et~al.}(2018)\citenamefont {Paneru},
  \citenamefont {Lee}, \citenamefont {Tlusty},\ and\ \citenamefont
  {Pak}}]{Paneru2018}%
  \BibitemOpen
  \bibfield  {author} {\bibinfo {author} {\bibfnamefont {G.}~\bibnamefont
  {Paneru}}, \bibinfo {author} {\bibfnamefont {D.~Y.}\ \bibnamefont {Lee}},
  \bibinfo {author} {\bibfnamefont {T.}~\bibnamefont {Tlusty}},\ and\ \bibinfo
  {author} {\bibfnamefont {H.~K.}\ \bibnamefont {Pak}},\ }\bibfield  {title}
  {\bibinfo {title} {{Lossless Brownian Information Engine}},\ }\href
  {https://doi.org/10.1103/PhysRevLett.120.020601} {\bibfield  {journal}
  {\bibinfo  {journal} {Physical Review Letters}\ }\textbf {\bibinfo {volume}
  {120}},\ \bibinfo {pages} {020601} (\bibinfo {year} {2018})},\ \Eprint
  {https://arxiv.org/abs/1802.01868} {arXiv:1802.01868} \BibitemShut {NoStop}%
\bibitem [{\citenamefont {Talkner}\ \emph {et~al.}(2007)\citenamefont
  {Talkner}, \citenamefont {Lutz},\ and\ \citenamefont
  {H{\"{a}}nggi}}]{Talkner2007}%
  \BibitemOpen
  \bibfield  {author} {\bibinfo {author} {\bibfnamefont {P.}~\bibnamefont
  {Talkner}}, \bibinfo {author} {\bibfnamefont {E.}~\bibnamefont {Lutz}},\ and\
  \bibinfo {author} {\bibfnamefont {P.}~\bibnamefont {H{\"{a}}nggi}},\
  }\bibfield  {title} {\bibinfo {title} {{Fluctuation theorems: Work is not an
  observable}},\ }\href {https://doi.org/10.1103/PhysRevE.75.050102} {\bibfield
   {journal} {\bibinfo  {journal} {Physical Review E - Statistical, Nonlinear,
  and Soft Matter Physics}\ }\textbf {\bibinfo {volume} {75}},\ \bibinfo
  {pages} {050102} (\bibinfo {year} {2007})},\ \Eprint
  {https://arxiv.org/abs/cond-mat/0703189} {arXiv:cond-mat/0703189}
  \BibitemShut {NoStop}%
\bibitem [{\citenamefont {Paik}\ \emph {et~al.}(2011)\citenamefont {Paik},
  \citenamefont {Schuster}, \citenamefont {Bishop}, \citenamefont {Kirchmair},
  \citenamefont {Catelani}, \citenamefont {Sears}, \citenamefont {Johnson},
  \citenamefont {Reagor}, \citenamefont {Frunzio}, \citenamefont {Glazman},
  \citenamefont {Girvin}, \citenamefont {Devoret},\ and\ \citenamefont
  {Schoelkopf}}]{Paik2011}%
  \BibitemOpen
  \bibfield  {author} {\bibinfo {author} {\bibfnamefont {H.}~\bibnamefont
  {Paik}}, \bibinfo {author} {\bibfnamefont {D.~I.}\ \bibnamefont {Schuster}},
  \bibinfo {author} {\bibfnamefont {L.~S.}\ \bibnamefont {Bishop}}, \bibinfo
  {author} {\bibfnamefont {G.}~\bibnamefont {Kirchmair}}, \bibinfo {author}
  {\bibfnamefont {G.}~\bibnamefont {Catelani}}, \bibinfo {author}
  {\bibfnamefont {A.~P.}\ \bibnamefont {Sears}}, \bibinfo {author}
  {\bibfnamefont {B.~R.}\ \bibnamefont {Johnson}}, \bibinfo {author}
  {\bibfnamefont {M.~J.}\ \bibnamefont {Reagor}}, \bibinfo {author}
  {\bibfnamefont {L.}~\bibnamefont {Frunzio}}, \bibinfo {author} {\bibfnamefont
  {L.~I.}\ \bibnamefont {Glazman}}, \bibinfo {author} {\bibfnamefont {S.~M.}\
  \bibnamefont {Girvin}}, \bibinfo {author} {\bibfnamefont {M.~H.}\
  \bibnamefont {Devoret}},\ and\ \bibinfo {author} {\bibfnamefont {R.~J.}\
  \bibnamefont {Schoelkopf}},\ }\bibfield  {title} {\bibinfo {title}
  {{Observation of high coherence in Josephson junction qubits measured in a
  three-dimensional circuit QED architecture}},\ }\href
  {https://doi.org/10.1103/PhysRevLett.107.240501} {\bibfield  {journal}
  {\bibinfo  {journal} {Physical Review Letters}\ }\textbf {\bibinfo {volume}
  {107}},\ \bibinfo {pages} {1} (\bibinfo {year} {2011})},\ \Eprint
  {https://arxiv.org/abs/1105.4652} {arXiv:1105.4652} \BibitemShut {NoStop}%
\bibitem [{\citenamefont {Koch}\ \emph {et~al.}(2007)\citenamefont {Koch},
  \citenamefont {Yu}, \citenamefont {Gambetta}, \citenamefont {Houck},
  \citenamefont {Schuster}, \citenamefont {Majer}, \citenamefont {Blais},
  \citenamefont {Devoret}, \citenamefont {Girvin},\ and\ \citenamefont
  {Schoelkopf}}]{Koch2007}%
  \BibitemOpen
  \bibfield  {author} {\bibinfo {author} {\bibfnamefont {J.}~\bibnamefont
  {Koch}}, \bibinfo {author} {\bibfnamefont {T.~M.}\ \bibnamefont {Yu}},
  \bibinfo {author} {\bibfnamefont {J.}~\bibnamefont {Gambetta}}, \bibinfo
  {author} {\bibfnamefont {A.~A.}\ \bibnamefont {Houck}}, \bibinfo {author}
  {\bibfnamefont {D.~I.}\ \bibnamefont {Schuster}}, \bibinfo {author}
  {\bibfnamefont {J.}~\bibnamefont {Majer}}, \bibinfo {author} {\bibfnamefont
  {A.}~\bibnamefont {Blais}}, \bibinfo {author} {\bibfnamefont {M.~H.}\
  \bibnamefont {Devoret}}, \bibinfo {author} {\bibfnamefont {S.~M.}\
  \bibnamefont {Girvin}},\ and\ \bibinfo {author} {\bibfnamefont {R.~J.}\
  \bibnamefont {Schoelkopf}},\ }\bibfield  {title} {\bibinfo {title}
  {{Charge-insensitive qubit design derived from the Cooper pair box}},\ }\href
  {https://doi.org/10.1103/PhysRevA.76.042319} {\bibfield  {journal} {\bibinfo
  {journal} {Physical Review A - Atomic, Molecular, and Optical Physics}\
  }\textbf {\bibinfo {volume} {76}},\ \bibinfo {pages} {042319} (\bibinfo
  {year} {2007})},\ \Eprint {https://arxiv.org/abs/cond-mat/0703002}
  {arXiv:cond-mat/0703002} \BibitemShut {NoStop}%
\bibitem [{\citenamefont {Wallraff}\ \emph {et~al.}(2004)\citenamefont
  {Wallraff}, \citenamefont {Schuster}, \citenamefont {Blais}, \citenamefont
  {Frunzio}, \citenamefont {Huang}, \citenamefont {Majer}, \citenamefont
  {Kumar}, \citenamefont {Girvin},\ and\ \citenamefont
  {Schoelkopf}}]{Wallraff2004}%
  \BibitemOpen
  \bibfield  {author} {\bibinfo {author} {\bibfnamefont {A.}~\bibnamefont
  {Wallraff}}, \bibinfo {author} {\bibfnamefont {D.~I.}\ \bibnamefont
  {Schuster}}, \bibinfo {author} {\bibfnamefont {A.}~\bibnamefont {Blais}},
  \bibinfo {author} {\bibfnamefont {L.}~\bibnamefont {Frunzio}}, \bibinfo
  {author} {\bibfnamefont {R.~S.}\ \bibnamefont {Huang}}, \bibinfo {author}
  {\bibfnamefont {J.}~\bibnamefont {Majer}}, \bibinfo {author} {\bibfnamefont
  {S.}~\bibnamefont {Kumar}}, \bibinfo {author} {\bibfnamefont {S.~M.}\
  \bibnamefont {Girvin}},\ and\ \bibinfo {author} {\bibfnamefont {R.~J.}\
  \bibnamefont {Schoelkopf}},\ }\bibfield  {title} {\bibinfo {title} {{Strong
  coupling of a single photon to a superconducting qubit using circuit quantum
  electrodynamics}},\ }\href {https://doi.org/10.1038/nature02851} {\bibfield
  {journal} {\bibinfo  {journal} {Nature}\ }\textbf {\bibinfo {volume} {431}},\
  \bibinfo {pages} {162} (\bibinfo {year} {2004})}\BibitemShut {NoStop}%
\bibitem [{\citenamefont {Hatridge}\ \emph {et~al.}(2013)\citenamefont
  {Hatridge}, \citenamefont {Shankar}, \citenamefont {Mirrahimi}, \citenamefont
  {Schackert}, \citenamefont {Geerlings}, \citenamefont {Brecht}, \citenamefont
  {Sliwa}, \citenamefont {Abdo}, \citenamefont {Frunzio}, \citenamefont
  {Girvin}, \citenamefont {Schoelkopf},\ and\ \citenamefont
  {Devoret}}]{Hatridge2013}%
  \BibitemOpen
  \bibfield  {author} {\bibinfo {author} {\bibfnamefont {M.}~\bibnamefont
  {Hatridge}}, \bibinfo {author} {\bibfnamefont {S.}~\bibnamefont {Shankar}},
  \bibinfo {author} {\bibfnamefont {M.}~\bibnamefont {Mirrahimi}}, \bibinfo
  {author} {\bibfnamefont {F.}~\bibnamefont {Schackert}}, \bibinfo {author}
  {\bibfnamefont {K.}~\bibnamefont {Geerlings}}, \bibinfo {author}
  {\bibfnamefont {T.}~\bibnamefont {Brecht}}, \bibinfo {author} {\bibfnamefont
  {K.~M.}\ \bibnamefont {Sliwa}}, \bibinfo {author} {\bibfnamefont
  {B.}~\bibnamefont {Abdo}}, \bibinfo {author} {\bibfnamefont {L.}~\bibnamefont
  {Frunzio}}, \bibinfo {author} {\bibfnamefont {S.~M.}\ \bibnamefont {Girvin}},
  \bibinfo {author} {\bibfnamefont {R.~J.}\ \bibnamefont {Schoelkopf}},\ and\
  \bibinfo {author} {\bibfnamefont {M.~H.}\ \bibnamefont {Devoret}},\
  }\bibfield  {title} {\bibinfo {title} {{Quantum back-action of an individual
  variable-strength measurement}},\ }\href
  {https://doi.org/10.1126/science.1226897} {\bibfield  {journal} {\bibinfo
  {journal} {Science}\ }\textbf {\bibinfo {volume} {339}},\ \bibinfo {pages}
  {178} (\bibinfo {year} {2013})},\ \Eprint {https://arxiv.org/abs/1903.11732}
  {arXiv:1903.11732} \BibitemShut {NoStop}%
\bibitem [{\citenamefont {Murch}\ \emph {et~al.}(2013)\citenamefont {Murch},
  \citenamefont {Weber}, \citenamefont {Macklin},\ and\ \citenamefont
  {Siddiqi}}]{Murch2013}%
  \BibitemOpen
  \bibfield  {author} {\bibinfo {author} {\bibfnamefont {K.~W.}\ \bibnamefont
  {Murch}}, \bibinfo {author} {\bibfnamefont {S.~J.}\ \bibnamefont {Weber}},
  \bibinfo {author} {\bibfnamefont {C.}~\bibnamefont {Macklin}},\ and\ \bibinfo
  {author} {\bibfnamefont {I.}~\bibnamefont {Siddiqi}},\ }\bibfield  {title}
  {\bibinfo {title} {{Observing single quantum trajectories of a
  superconducting quantum bit}},\ }\href {https://doi.org/10.1038/nature12539}
  {\bibfield  {journal} {\bibinfo  {journal} {Nature}\ }\textbf {\bibinfo
  {volume} {502}},\ \bibinfo {pages} {211} (\bibinfo {year} {2013})},\ \Eprint
  {https://arxiv.org/abs/1305.7270} {arXiv:1305.7270} \BibitemShut {NoStop}%
\bibitem [{\citenamefont {Weber}\ \emph {et~al.}(2014)\citenamefont {Weber},
  \citenamefont {Chantasri}, \citenamefont {Dressel}, \citenamefont {Jordan},
  \citenamefont {Murch},\ and\ \citenamefont {Siddiqi}}]{Weber2014}%
  \BibitemOpen
  \bibfield  {author} {\bibinfo {author} {\bibfnamefont {S.~J.}\ \bibnamefont
  {Weber}}, \bibinfo {author} {\bibfnamefont {A.}~\bibnamefont {Chantasri}},
  \bibinfo {author} {\bibfnamefont {J.}~\bibnamefont {Dressel}}, \bibinfo
  {author} {\bibfnamefont {A.~N.}\ \bibnamefont {Jordan}}, \bibinfo {author}
  {\bibfnamefont {K.~W.}\ \bibnamefont {Murch}},\ and\ \bibinfo {author}
  {\bibfnamefont {I.}~\bibnamefont {Siddiqi}},\ }\bibfield  {title} {\bibinfo
  {title} {{Mapping the optimal route between two quantum states}},\ }\href
  {https://doi.org/10.1038/nature13559} {\bibfield  {journal} {\bibinfo
  {journal} {Nature}\ }\textbf {\bibinfo {volume} {511}},\ \bibinfo {pages}
  {570} (\bibinfo {year} {2014})},\ \Eprint {https://arxiv.org/abs/1403.4992}
  {arXiv:1403.4992} \BibitemShut {NoStop}%
\bibitem [{\citenamefont {Gambetta}\ \emph {et~al.}(2008)\citenamefont
  {Gambetta}, \citenamefont {Blais}, \citenamefont {Boissonneault},
  \citenamefont {Houck}, \citenamefont {Schuster},\ and\ \citenamefont
  {Girvin}}]{Gambetta2008}%
  \BibitemOpen
  \bibfield  {author} {\bibinfo {author} {\bibfnamefont {J.}~\bibnamefont
  {Gambetta}}, \bibinfo {author} {\bibfnamefont {A.}~\bibnamefont {Blais}},
  \bibinfo {author} {\bibfnamefont {M.}~\bibnamefont {Boissonneault}}, \bibinfo
  {author} {\bibfnamefont {A.~A.}\ \bibnamefont {Houck}}, \bibinfo {author}
  {\bibfnamefont {D.~I.}\ \bibnamefont {Schuster}},\ and\ \bibinfo {author}
  {\bibfnamefont {S.~M.}\ \bibnamefont {Girvin}},\ }\bibfield  {title}
  {\bibinfo {title} {{Quantum trajectory approach to circuit QED: Quantum jumps
  and the Zeno effect}},\ }\href {https://doi.org/10.1103/PhysRevA.77.012112}
  {\bibfield  {journal} {\bibinfo  {journal} {Physical Review A - Atomic,
  Molecular, and Optical Physics}\ }\textbf {\bibinfo {volume} {77}},\ \bibinfo
  {pages} {1} (\bibinfo {year} {2008})},\ \Eprint
  {https://arxiv.org/abs/0709.4264} {arXiv:0709.4264} \BibitemShut {NoStop}%
\bibitem [{\citenamefont {Tan}\ \emph {et~al.}(2014)\citenamefont {Tan},
  \citenamefont {Weber}, \citenamefont {Siddiqi}, \citenamefont {M{\o}lmer},\
  and\ \citenamefont {Murch}}]{Tan2015}%
  \BibitemOpen
  \bibfield  {author} {\bibinfo {author} {\bibfnamefont {D.}~\bibnamefont
  {Tan}}, \bibinfo {author} {\bibfnamefont {S.}~\bibnamefont {Weber}}, \bibinfo
  {author} {\bibfnamefont {I.}~\bibnamefont {Siddiqi}}, \bibinfo {author}
  {\bibfnamefont {K.}~\bibnamefont {M{\o}lmer}},\ and\ \bibinfo {author}
  {\bibfnamefont {K.~W.}\ \bibnamefont {Murch}},\ }\bibfield  {title} {\bibinfo
  {title} {{Prediction and retrodiction for a continuously monitored
  superconducting qubit}},\ }\href
  {https://doi.org/10.1103/PhysRevLett.114.090403} {\bibfield  {journal}
  {\bibinfo  {journal} {Physical Review Letters}\ }\textbf {\bibinfo {volume}
  {114}},\ \bibinfo {pages} {090403} (\bibinfo {year} {2014})},\ \Eprint
  {https://arxiv.org/abs/1409.0510} {arXiv:1409.0510} \BibitemShut {NoStop}%
\bibitem [{\citenamefont {Foroozani}\ \emph {et~al.}(2016)\citenamefont
  {Foroozani}, \citenamefont {Naghiloo}, \citenamefont {Tan}, \citenamefont
  {M{\o}lmer},\ and\ \citenamefont {Murch}}]{Foroozani2016}%
  \BibitemOpen
  \bibfield  {author} {\bibinfo {author} {\bibfnamefont {N.}~\bibnamefont
  {Foroozani}}, \bibinfo {author} {\bibfnamefont {M.}~\bibnamefont {Naghiloo}},
  \bibinfo {author} {\bibfnamefont {D.}~\bibnamefont {Tan}}, \bibinfo {author}
  {\bibfnamefont {K.}~\bibnamefont {M{\o}lmer}},\ and\ \bibinfo {author}
  {\bibfnamefont {K.~W.}\ \bibnamefont {Murch}},\ }\bibfield  {title} {\bibinfo
  {title} {{Correlations of the Time Dependent Signal and the State of a
  Continuously Monitored Quantum System}},\ }\href
  {https://doi.org/10.1103/PhysRevLett.116.110401} {\bibfield  {journal}
  {\bibinfo  {journal} {Physical Review Letters}\ }\textbf {\bibinfo {volume}
  {116}},\ \bibinfo {pages} {110401} (\bibinfo {year} {2016})},\ \Eprint
  {https://arxiv.org/abs/1508.01185} {arXiv:1508.01185} \BibitemShut {NoStop}%
\bibitem [{\citenamefont {Jacobs}(2006)}]{Jacobs2006}%
  \BibitemOpen
  \bibfield  {author} {\bibinfo {author} {\bibfnamefont {K.}~\bibnamefont
  {Jacobs}},\ }\bibfield  {title} {\bibinfo {title} {{A bound on the mutual
  information, and properties of entropy reduction, for quantum channels with
  inefficient measurements}},\ }\href {https://doi.org/10.1063/1.2158433}
  {\bibfield  {journal} {\bibinfo  {journal} {Journal of Mathematical Physics}\
  }\textbf {\bibinfo {volume} {47}},\ \bibinfo {pages} {012102} (\bibinfo
  {year} {2006})},\ \Eprint {https://arxiv.org/abs/quant-ph/0412006}
  {arXiv:quant-ph/0412006} \BibitemShut {NoStop}%
\bibitem [{\citenamefont {Nielsen}\ and\ \citenamefont
  {Chuang}(2010)}]{Nielsen2010}%
  \BibitemOpen
  \bibfield  {author} {\bibinfo {author} {\bibfnamefont {M.~A.}\ \bibnamefont
  {Nielsen}}\ and\ \bibinfo {author} {\bibfnamefont {I.~L.}\ \bibnamefont
  {Chuang}},\ }\href {https://doi.org/10.1017/cbo9780511976667} {\emph
  {\bibinfo {title} {Quantum Computation and Quantum Information}}}\ (\bibinfo
  {publisher} {Cambridge University Press},\ \bibinfo {address} {Cambridge},\
  \bibinfo {year} {2010})\BibitemShut {NoStop}%
\bibitem [{\citenamefont {Campisi}\ \emph {et~al.}(2011)\citenamefont
  {Campisi}, \citenamefont {H{\"{a}}nggi},\ and\ \citenamefont
  {Talkner}}]{Campisi2011}%
  \BibitemOpen
  \bibfield  {author} {\bibinfo {author} {\bibfnamefont {M.}~\bibnamefont
  {Campisi}}, \bibinfo {author} {\bibfnamefont {P.}~\bibnamefont
  {H{\"{a}}nggi}},\ and\ \bibinfo {author} {\bibfnamefont {P.}~\bibnamefont
  {Talkner}},\ }\bibfield  {title} {\bibinfo {title} {{Colloquium : Quantum
  fluctuation relations: Foundations and applications}},\ }\href
  {https://doi.org/10.1103/RevModPhys.83.771} {\bibfield  {journal} {\bibinfo
  {journal} {Reviews of Modern Physics}\ }\textbf {\bibinfo {volume} {83}},\
  \bibinfo {pages} {771} (\bibinfo {year} {2011})},\ \Eprint
  {https://arxiv.org/abs/1012.2268} {arXiv:1012.2268} \BibitemShut {NoStop}%
\bibitem [{\citenamefont {Jarzynski}(1997{\natexlab{a}})}]{Jarzynski1997}%
  \BibitemOpen
  \bibfield  {author} {\bibinfo {author} {\bibfnamefont {C.}~\bibnamefont
  {Jarzynski}},\ }\bibfield  {title} {\bibinfo {title} {{Nonequilibrium
  equality for free energy differences}},\ }\href
  {https://doi.org/10.1103/PhysRevLett.78.2690} {\bibfield  {journal} {\bibinfo
   {journal} {Physical Review Letters}\ }\textbf {\bibinfo {volume} {78}},\
  \bibinfo {pages} {2690} (\bibinfo {year} {1997}{\natexlab{a}})},\ \Eprint
  {https://arxiv.org/abs/cond-mat/9610209} {arXiv:cond-mat/9610209}
  \BibitemShut {NoStop}%
\bibitem [{\citenamefont {Jarzynski}(1997{\natexlab{b}})}]{Jarzynski1997a}%
  \BibitemOpen
  \bibfield  {author} {\bibinfo {author} {\bibfnamefont {C.}~\bibnamefont
  {Jarzynski}},\ }\bibfield  {title} {\bibinfo {title} {{Equilibrium
  free-energy differences from nonequilibrium measurements: A master-equation
  approach}},\ }\href {https://doi.org/10.1103/PhysRevE.56.5018} {\bibfield
  {journal} {\bibinfo  {journal} {Physical Review E - Statistical Physics,
  Plasmas, Fluids, and Related Interdisciplinary Topics}\ }\textbf {\bibinfo
  {volume} {56}},\ \bibinfo {pages} {5018} (\bibinfo {year}
  {1997}{\natexlab{b}})},\ \Eprint {https://arxiv.org/abs/cond-mat/9707325}
  {arXiv:cond-mat/9707325} \BibitemShut {NoStop}%
\bibitem [{\citenamefont {Sagawa}\ and\ \citenamefont
  {Ueda}(2010)}]{Sagawa2010}%
  \BibitemOpen
  \bibfield  {author} {\bibinfo {author} {\bibfnamefont {T.}~\bibnamefont
  {Sagawa}}\ and\ \bibinfo {author} {\bibfnamefont {M.}~\bibnamefont {Ueda}},\
  }\bibfield  {title} {\bibinfo {title} {{Generalized Jarzynski equality under
  nonequilibrium feedback control}},\ }\href
  {https://doi.org/10.1103/PhysRevLett.104.090602} {\bibfield  {journal}
  {\bibinfo  {journal} {Physical Review Letters}\ }\textbf {\bibinfo {volume}
  {104}},\ \bibinfo {pages} {090602} (\bibinfo {year} {2010})},\ \Eprint
  {https://arxiv.org/abs/0907.4914} {arXiv:0907.4914} \BibitemShut {NoStop}%
\bibitem [{\citenamefont {Crooks}(2008)}]{Crooks2008}%
  \BibitemOpen
  \bibfield  {author} {\bibinfo {author} {\bibfnamefont {G.~E.}\ \bibnamefont
  {Crooks}},\ }\bibfield  {title} {\bibinfo {title} {{Quantum operation time
  reversal}},\ }\href {https://doi.org/10.1103/physreva.77.034101} {\bibfield
  {journal} {\bibinfo  {journal} {Physical Review A}\ }\textbf {\bibinfo
  {volume} {77}},\ \bibinfo {pages} {1} (\bibinfo {year} {2008})},\ \Eprint
  {https://arxiv.org/abs/0706.3749} {arXiv:0706.3749} \BibitemShut {NoStop}%
\end{thebibliography}
%

\end{document}